%% file: main.tex
\documentclass[12pt,number,sort&compress]{elsarticle}

\usepackage{lineno}

\journal{Nucl. Instr. and Meth. A}

\newcommand{\gsim}{\hbox{ \raise3pt\hbox to 0pt{$>$}\raise-3pt\hbox{$\sim$} }}
\newcommand{\lsim}{\hbox{ \raise3pt\hbox to 0pt{$<$}\raise-3pt\hbox{$\sim$} }}
\newcommand{\del}{\ifmmode{\nabla}               \else{$\nabla$ }               \fi}

\newcommand{\figdir}{.}

\usepackage{comment}

\begin{document}

\begin{frontmatter}



\title{Cosmic ray tests of a GEM-based TPC prototype \\
                   operated in Ar-CF$_4$-isobutane gas mixtures}
%
%
%
\input{authorlist.tex}
%
%
%
%
%
\begin{abstract}
Argon with an admixture of CF$_4$ is expected to be a good candidate for
the gas mixture to be used for a time projection chamber (TPC) in the
future linear collider experiment because of its small transverse
diffusion of drift electrons especially under a strong magnetic field.
In order to confirm the superiority of this gas mixture over conventional
TPC gases we carried out cosmic ray
tests using a GEM-based TPC operated mostly in Ar-CF$_4$-isobutane mixtures
under 0 - 1 T axial magnetic fields. The measured gas properties such as
gas gain and transverse diffusion constant as well as the observed
spatial resolution are presented.
\end{abstract}
%
%
%
%
\begin{keyword}


TPC\sep
ILC\sep
GEM\sep
CF$_4$\sep
Diffusion\sep
Spatial resolution\sep



\PACS
29.40.Cs \sep
29.40.Gx

\end{keyword}

\end{frontmatter}


%
%
%
%
%
%

\section{Introduction}

A strong candidate for the central tracker of the future
linear collider (LC) experiment~\cite{LC1,LC2}
is a large time projection chamber (TPC)~\cite{Nygren}.
The LCTPC is expected to have unprecedentedly high momentum resolution
under a 3 - 4 T magnetic field and good two-track resolving power to
precisely reconstruct high momentum lepton tracks and each of the
charged tracks in densely packed jets.
In order to fulfill these requirements a micro-pattern gas detector
(MPGD) is planned to be used for the endplate readout device
since a conventional multi-wire proportional chamber (MWPC) readout
suffers severely from the $E \times B$ effect in the vicinity of its sense
wire planes under a high magnetic field~\cite{arXiv1}.
A GEM-based TPC prototype has been proved to work stably in
conventional gas mixtures and to give satisfactory spatial resolution
(see, for example, Ref.~\cite{Kobayashi2}).

However, the target spatial resolution of better than 100 $\mu$m in the
$r$-$\phi$ plane for high momentum tracks is still ambitious and difficult
to achieve with
conventional gas mixtures such as TDR gas (Ar-CH$_4$(5\%)-CO$_2$(2\%))
or P5 gas (Ar-CH$_4$(5\%)), in particular at long drift distances
($\ge$ 2 m) because of unavoidable diffusion of drift electrons
even under a strong axial magnetic field.

The spatial resolution in the pad row direction ($\sigma_{\rm X}$)
as a function of the drift distance ($z$) is expressed as 
\begin{equation}
\sigma_{\rm X}^2 = \sigma_{\rm X0}^2 + \frac{D^2}{N_{\rm eff}} \cdot z \;,
\end{equation}
where $\sigma_{\rm X0}$ is a constant term, $D$ is the transverse
diffusion constant\footnote
{
The diffusion constant $D$ is related to the diffusion coefficient ($D^*$) through
$D^2 = 2D^*/W$, where $W$ is the electron drift velocity.  
},
and $N_{\rm eff}$ the effective number of
electrons\footnote
{
Equation~(1) may be inappropriate for small $z$ as will be seen in Section~4.2.
Even in that case, however, it describes the asymptotic behavior of the 
spatial resolution at long drift distances. 
}.
It should be noted that Eq.~(1) contains no angular terms (the angular
wire and the $E \times B$ effects) except for the angular pad effect, which
is implicitly included in the constant term ($\sigma_{\rm X0}$).
The value of $\sigma_{\rm X0}$ becomes large as the track angle increases with respect
to the pad-row normal.

\begin{figure*}[t]
\begin{center}
\hspace{10mm}
\includegraphics*[scale=0.55]{\figdir/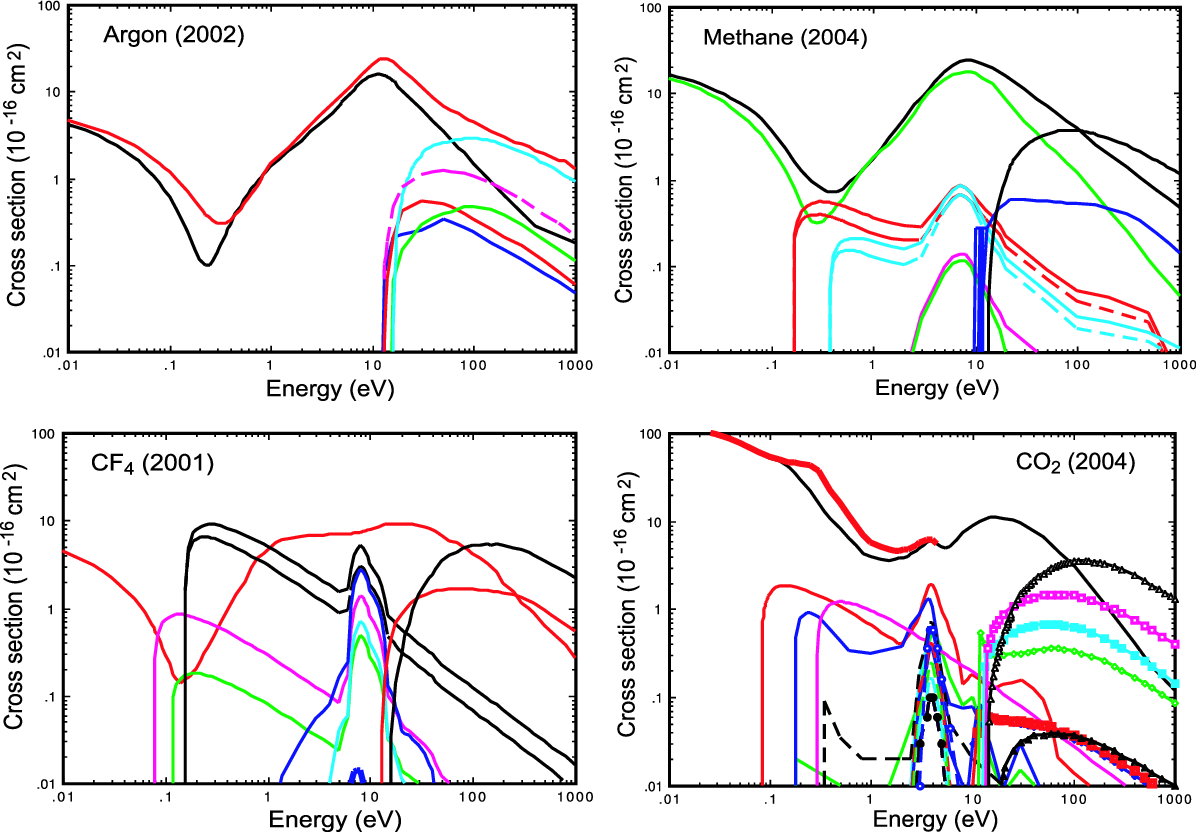}
\end{center}
\caption{\label{fig1}
\footnotesize Cross sections of argon, CH$_4$, CF$_4$ and CO$_2$
as function of the electron energy~\cite{Magboltz-writeup}.
See Refs.~\cite{Magboltz-writeup,Biagi} for each of the components contributing
to the cross sections shown in curves.
}
\end{figure*}

From our experience $\sigma_{\rm X0}$ is about 50 $\mu$m (or better)
for tracks perpendicular to the pad row direction
if the charge spread in the amplification gap is not too small compared to the
readout pad pitch and the S/N ratio of the readout electronics is good enough.
The parameter $N_{\rm eff}$ has been measured to be
around 20 in argon-based gases and for a pad height of
$\sim$ 6 mm~\cite{Kobayashi2,Paul},
which is consistent with a simple estimate taking into account the
primary ionization statistics and the avalanche fluctuation in the
amplification gap~\cite{Kobayashi}.  

If we require the azimuthal resolution of 100 $\mu$m at z = 200 cm
the diffusion constant ($D$), which is essentially the only free
(controllable) parameter depending on the choice of gas mixture,
needs to be smaller than 30 $\mu$m/$\sqrt{\rm cm}$.
The diffusion constant of drift electrons under the influence of an
axial magnetic field ($B$) is given by
\begin{equation}
D(B) = D(B = 0)/\sqrt{1 + (\omega \tau)^2} \;,
\end{equation}
where $\omega \equiv e \cdot B/m$, the electron cyclotron frequency, and $\tau$
is the mean free time of drift electrons between collisions with gas molecules.
Therefore we need a gas mixture in which $D(B=0)$ is small ({\it cool\/}) and
$\tau$ is fairly large ({\it fast\/}) under a moderate drift field
($E$).\footnote
{
The electron drift velocity is given by $W = e \cdot E/m \cdot \tau$
with $e$ ($m$) being the electron charge (mass).
A large value of $\tau$, therefore, means a fast gas.
}

A guideline to obtain a cool and fast gas is suggested by
the deep and broad
Ramsauer minimum in the electron cross section in argon,
which is located around the electron energy ($\epsilon$) of 0.2 eV.
It is necessary somehow to confine drift electrons near the Ramsauer
minimum in argon, where the electrons are relatively cool and the cross
section is small ($\tau$ is large), under a moderate drift field.
Actually, argon itself is a very slow and hot gas and the drift electrons get
hot quickly even under a weak drift field.
Therefore it is necessary to add an efficient moderator, i.e. a small amount of
molecular gas with its Ramsauer minimum at around $\epsilon$ = 0.2 eV and with
large inelastic cross sections at $\epsilon \ge$ 0.2 eV in order to
efficiently cool down relatively hot electrons~\cite{Christophorou1,Christophorou2}.

\begin{figure}[htbp]
\begin{center}
\hspace{10mm}
\includegraphics*[scale=0.35]{\figdir/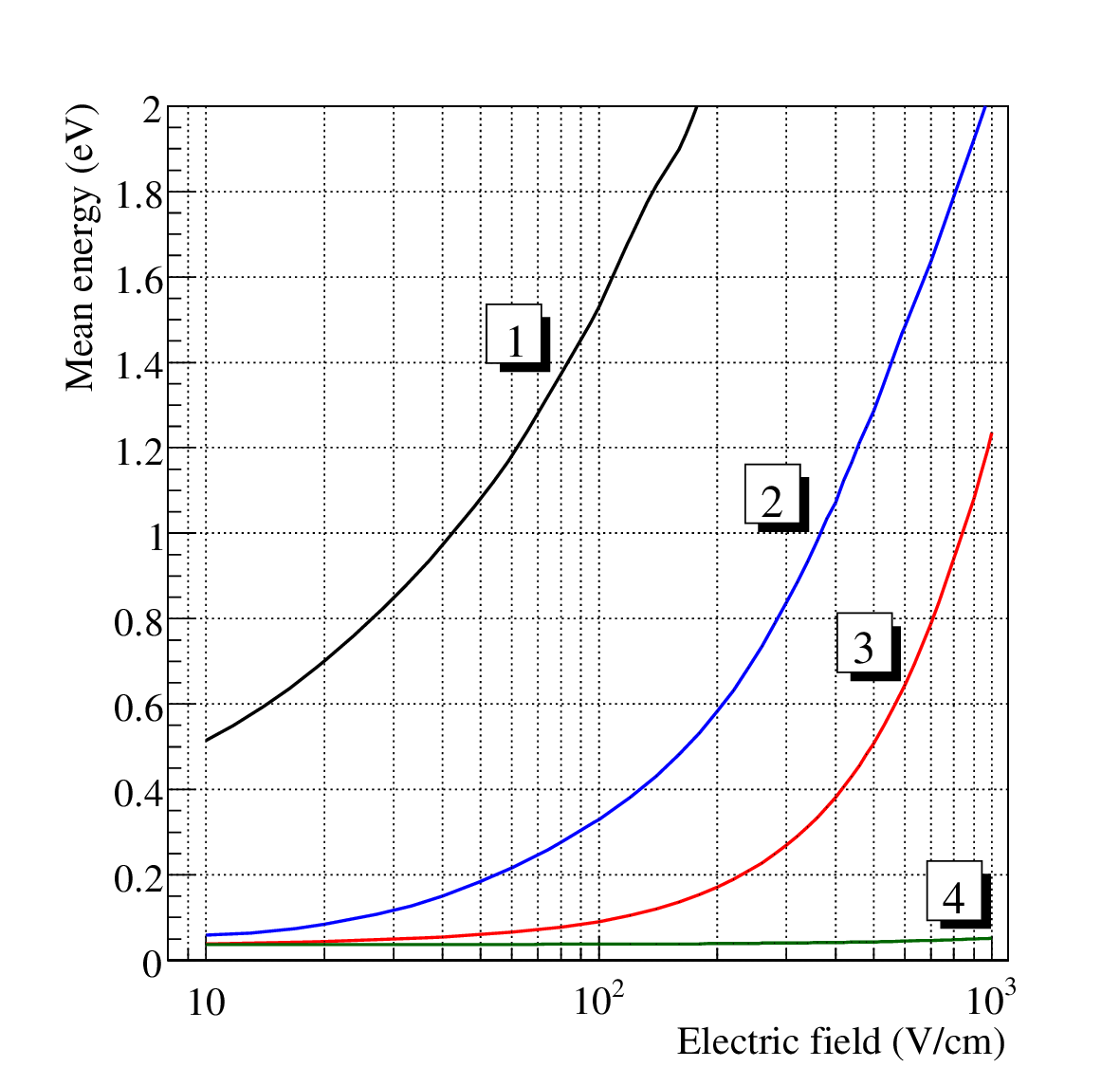}
\end{center}
\caption{\label{fig2}
\footnotesize Mean electron energy as a function of the electric field 
given by Magboltz for
(1) Ar (100\%), (2) Ar-CH$_4$(5\%), (3) Ar-CF$_4$(3\%)-isobutane(2\%),
and (4) CO$_2$ (100\%).
Electrons are almost thermal in pure CO$_2$ in the range of electric field shown
in the figure. 
}
\end{figure}

\begin{figure}[htbp]
\begin{center}
\hspace{10mm}
\includegraphics*[scale=0.42]{\figdir/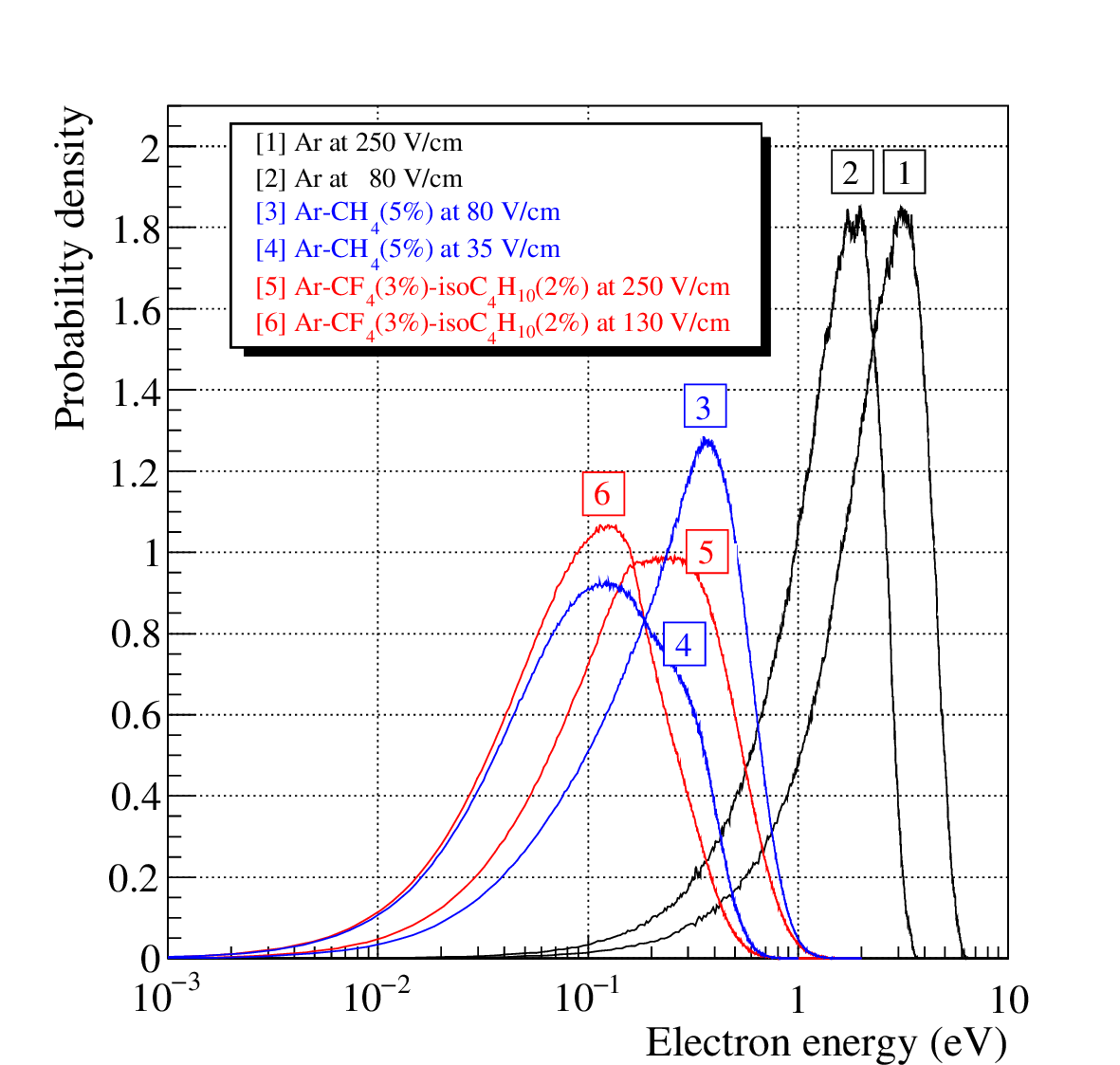}
\end{center}
\caption{\label{fig3}
\footnotesize Electron energy distribution in some gases given by Magboltz.
The electric field of 80 V/cm (250 V/cm) approximately corresponds to the
drift-velocity maximum for Ar-CH$_4$(5\%) (Ar-CF$_4$(3\%)-isobutane(2\%)) while
the electric field of 35 V/cm (130 V/cm) corresponds to the diffusion minimum 
for Ar-CH$_4$(5\%) (Ar-CF$_4$(3\%)-isobutane(2\%)).
}
\end{figure}

Good candidates for the additive gas are CH$_4$ and CF$_4$.
The cross sections of relevant gases to electrons are seen in Fig.~\ref{fig1}
as function of the electron energy.
Figure~\ref{fig2} shows the mean electron energy as a function of the electric field
for some pure gases and gas mixtures given by Magboltz~\cite{Biagi}\footnote{
The mean electron energy can be estimated experimentally from the characteristic energy
($\epsilon_{\rm k}$): $\left< \epsilon \right> \approx \frac{3}{2} \epsilon_{\rm k}
\equiv \frac{3}{2} \cdot \frac{e \cdot D^*}{\mu}$, where $D^*$ is the diffusion coefficient
and $\mu$ the mobility of electrons \cite{Huxley}. 
}.
All the results of Magboltz (version 8.5) presented in the present work are for
gases at NTP (20$^{\circ}$C, 1 atm).
Examples of the electron energy distribution are shown in Fig.~\ref{fig3} for 
some gases of interest at several electric fields.
The figure demonstrates that CH$_4$ or CF$_4$ molecules in argon are very efficient to
keep the electron temperatures around those corresponding to the Ramsauer minimum,
even under relatively high electric fields.
Furthermore, CF$_4$ is superior to CH$_4$ as a moderator additive because of its
much larger inelastic cross section to electrons with energies above the
Ramsauer dip, due to vibrational excitation (see Fig. 1).


Although they are fast and (relatively) cool themselves it is favorable to
dilute them with argon in order to lower the drift field corresponding
to the drift-velocity maximum by effectively reducing their inelastic
cross sections, and at the same time, the operating high voltage of MPGDs.
A low drift field is favorable especially for a large TPC
in order to limit the high voltage for the central membrane to a moderate level.
It is worth mentioning here that the drift field should be chosen
so that the drift velocity be as close as possible to its maximum
since the gas pressure in the LCTPC follows atmospheric pressure.

We have already tested several candidate gases, including TDR gas and P5 gas,
for the LCTPC using a small prototype equipped with an MPGD
(MicroMEGAS~\cite{Giomataris} or GEM~\cite{Sauli})
in beam tests~\cite{Kobayashi2,Paul}.
The resultant transverse diffusion constants were always close to those
given by Magboltz under 0 - 1 T magnetic fields.
According to Magboltz the diffusion constant in P5 gas (TDR gas)
at $B$ = 4 T, and $E \sim$ 80 V/cm ($\sim$ 220 V/cm) corresponding to the
drift-velocity maximum,
is about 40 $\mu$m/$\sqrt{\rm cm}$ (60 $\mu$m/$\sqrt{\rm cm}$)~\cite{GasProperty},
which does not meet our requirement mentioned above.\footnote
{
A cool and {\it slow\/} gas such as DME (dimethyl ether) or CO$_2$ (as in the
case of TDR gas) is not a good additive since it makes the mean free
time of the drift electrons ($\tau$) small because of a large cross section
to low energy electrons (see Fig.~\ref{fig1} for the cross section of CO$_2$).
It certainly reduces the diffusion constant at $B = 0$ T but,
at the same time, makes the reduction of $D$ under a magnetic field
less effective.  
}

As for Ar-CF$_4$, which is cooler and faster than
P5 gas, we were not sure about the reliability of the Magboltz prediction
although some measurements using prototype TPCs mainly equipped with
MicroMEGAS
exist~\cite{CF4-papers-1,CF4-papers-2,CF4-papers-3,
CF4-papers-4,CF4-papers-5,CF4-papers-6,CF4-papers-7}.
We therefore conducted a series of cosmic ray tests using a small
prototype TPC with a GEM readout operated in this possibly promising
gas mixture of Ar-CF$_4$ (-isobutane) under a magnetic field of up to 1 T.
In addition to the measurement of diffusion constant it was important
as well to demonstrate successful operation of a GEM-equipped TPC in
gases containing CF$_4$ since CF$_4$ molecules dissociatively capture
relatively hot electrons in the vicinity of GEMs.\footnote
{
Significant deterioration of the energy resolution for soft X-rays due
to the electron attachment in gases containing CF$_4$ has been observed
with proportional tubes~\cite{Christophorou1,Thun,Anderson}
and with an MWPC-equipped TPC~\cite{Isobe}.
}
It should be noted that the loss of drift electrons at the entrance to
the first GEM could be fatal to the performance of a TPC, i.e. the spatial
and the d$E$/d$x$ resolutions, since this loss can not be compensated by
the subsequent gas multiplication.

The experimental setup is briefly described in Section~2.
The results of gas gain measurements are given in Section~3,
while the measured drift properties of the gas and the performance of the
prototype are presented in Section~4.
Section~5 is devoted to the conclusion.

\section{Experimental setup}

\begin{figure}[htbp]
\begin{center}
\hspace{10mm}
\includegraphics*[scale=0.525]{\figdir/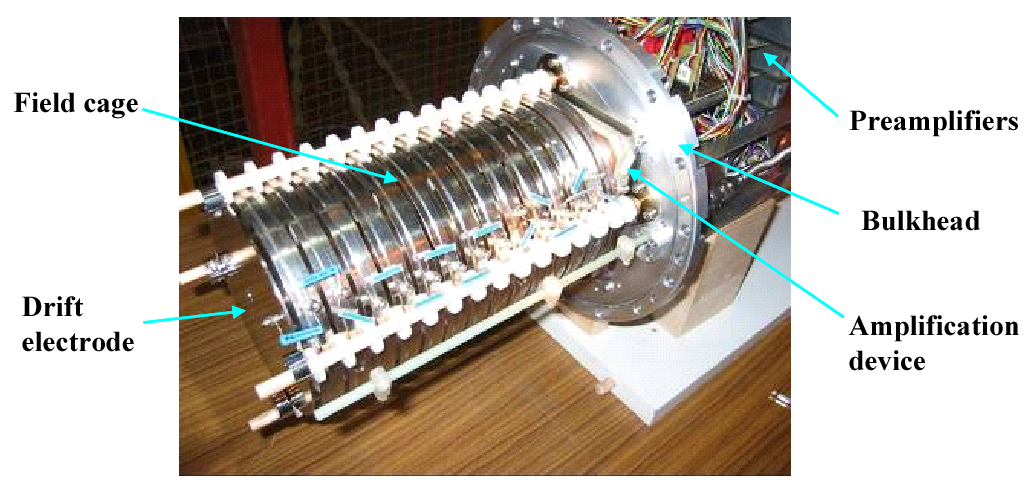}
\end{center}
\caption{\label{fig4}
\footnotesize Photograph of the prototype just before installation
into the gas vessel.
}
\end{figure}

A photograph of the prototype (MP-TPC) is shown in Fig.~\ref{fig4}.
It consists of a field cage and an easily replaceable gas amplification
device attached to one end of the field cage.
Gas amplified electrons are detected by a pad plane at ground potential
placed right behind the amplification device.
A drift electrode is attached to the other end of the field cage.
The maximum drift length is 257 mm.

In the present test a triple GEM is used as the amplification device.
The triple GEM, CERN standard~\cite{Bachmann}, has 1.5 mm transfer gaps and a
1.5 mm induction gap.
The high voltages applied across each of the GEM foils
and the transfer and induction gaps
are set equal with a resistor-chain voltage divider. 

The pad plane, with an effective area of $100 \times 100$ mm$^2$,
has 16 pad rows at a pitch of 6.3 mm, each consisting of
$1.17 \times 6$ mm$^2$ rectangular pads arranged at a pitch of 1.27 mm.
The neighboring pad rows are staggered by half a pad pitch.
Pad signals are fed to charge sensitive preamplifiers located 20 cm behind
the bulkhead of the gas vessel.
The amplified signals are sent to shaper amplifiers
via twisted pair cables, and then processed by 
12.5 MHz digitizers~\cite{arXiv2}.

The chamber gases are mainly Ar-CF$_4$-isobutane mixtures
at atmospheric pressure and room temperature.
The gas pressure and the ambient temperature ($\sim 20^\circ$C) are periodically
monitored since they are not controlled actively. 

\begin{figure}[htbp]
\begin{center}
\hspace{10mm}
\includegraphics*[scale=0.19]{\figdir/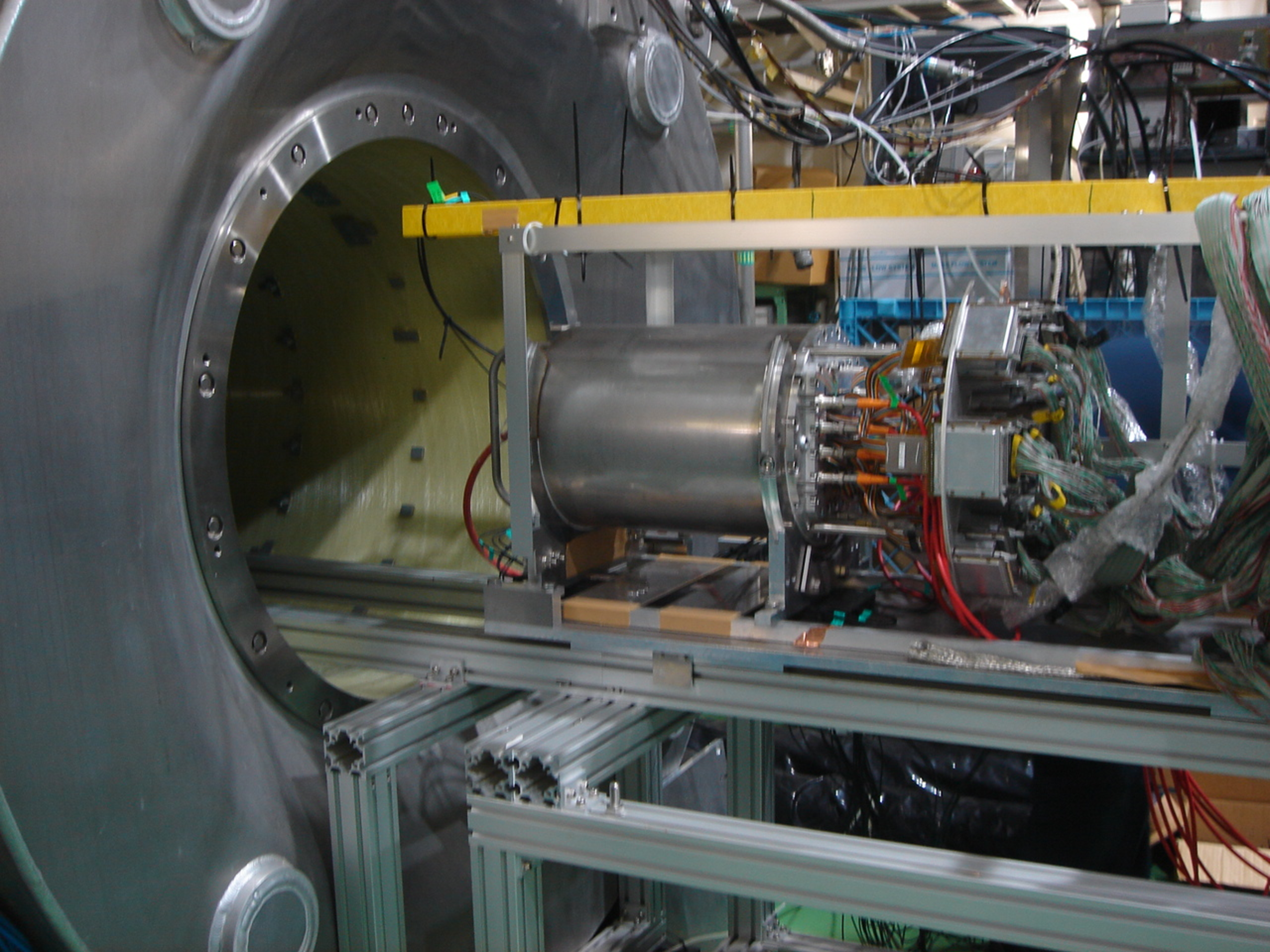}
\end{center}
\caption{\label{fig5}
\footnotesize Photograph of the prototype ready to be inserted into the magnet.
}
\end{figure}

The prototype TPC is placed in the uniform field region of a superconducting
solenoid without return yoke, having a bore diameter of 850 mm, an effective
length of 1000 mm, and a maximum field strength of 1.2 T (see Fig.~\ref{fig5}).
A pair of scintillation counters sandwiching the prototype TPC is used
to trigger a data acquisition system upon detection of cosmic ray tracks.

The measurement of gas gain was carried out using a dedicated small test
chamber and an $^{55}$Fe source without magnetic field.
The small chamber has a GEM stack with
a configuration similar to that of the prototype TPC
and a 10-mm drift gap for photon conversion
with a field strength of 200 V/cm.
Signal charge induced on its pad plane is read out by a combination of
a preamplifier, a shaper amplifier and an analog to digital converter.   

\section{Gas gain measurements}

A small value of the diffusion constant ($D\/$) and a reasonably large
effective number of electrons ($N_{\rm eff}$) are crucial properties 
of the LCTPC gas as mentioned in Section 1.
Another major practical issue is the gas gain in the amplification gap. 
The demonstration of stable operation of GEMs at an adequate gain is
important particularly in a gas containing CF$_4$ since CF$_4$ molecules 
dissociatively attach hot electrons during their passage through
the GEM stack. 
We therefore first measured the gas gains in several Ar-CF$_4$ based gas mixtures
as function of the high voltage across each GEM foil (V$_{\rm GEM}$).
The electric fields in the transfer and the induction gaps were kept at
1.6 kV/cm throughout the measurements. 

\begin{figure}[htbp]
\begin{center}
\hspace{10mm}
\includegraphics*[scale=0.330]{\figdir/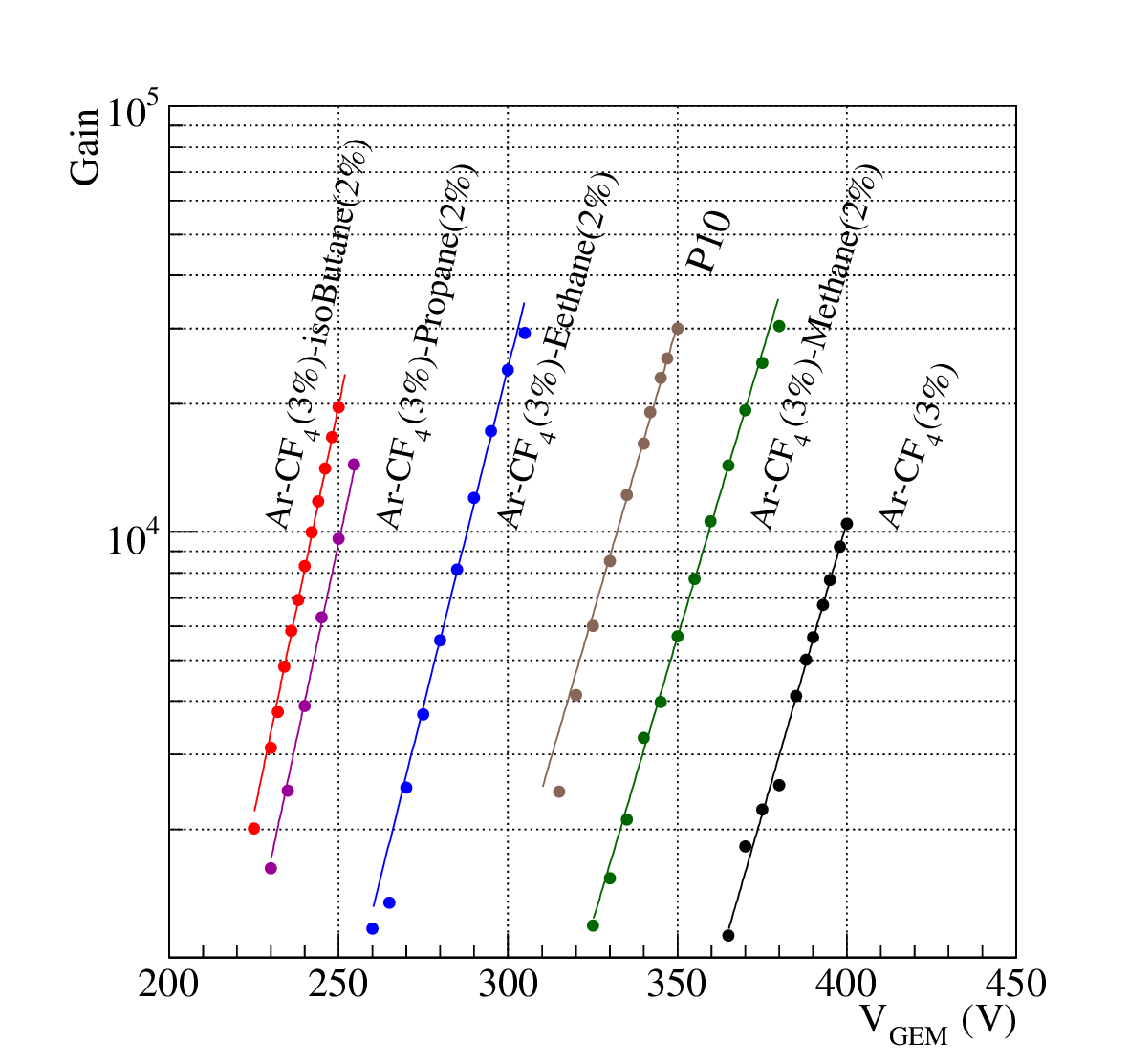}
\end{center}
\caption{\label{fig6}
\footnotesize Measured gas gain as a function of GEM Voltage.
}
\end{figure}

\begin{figure}[htbp]
\begin{center}
\hspace{10mm}
\includegraphics*[scale=0.28]{\figdir/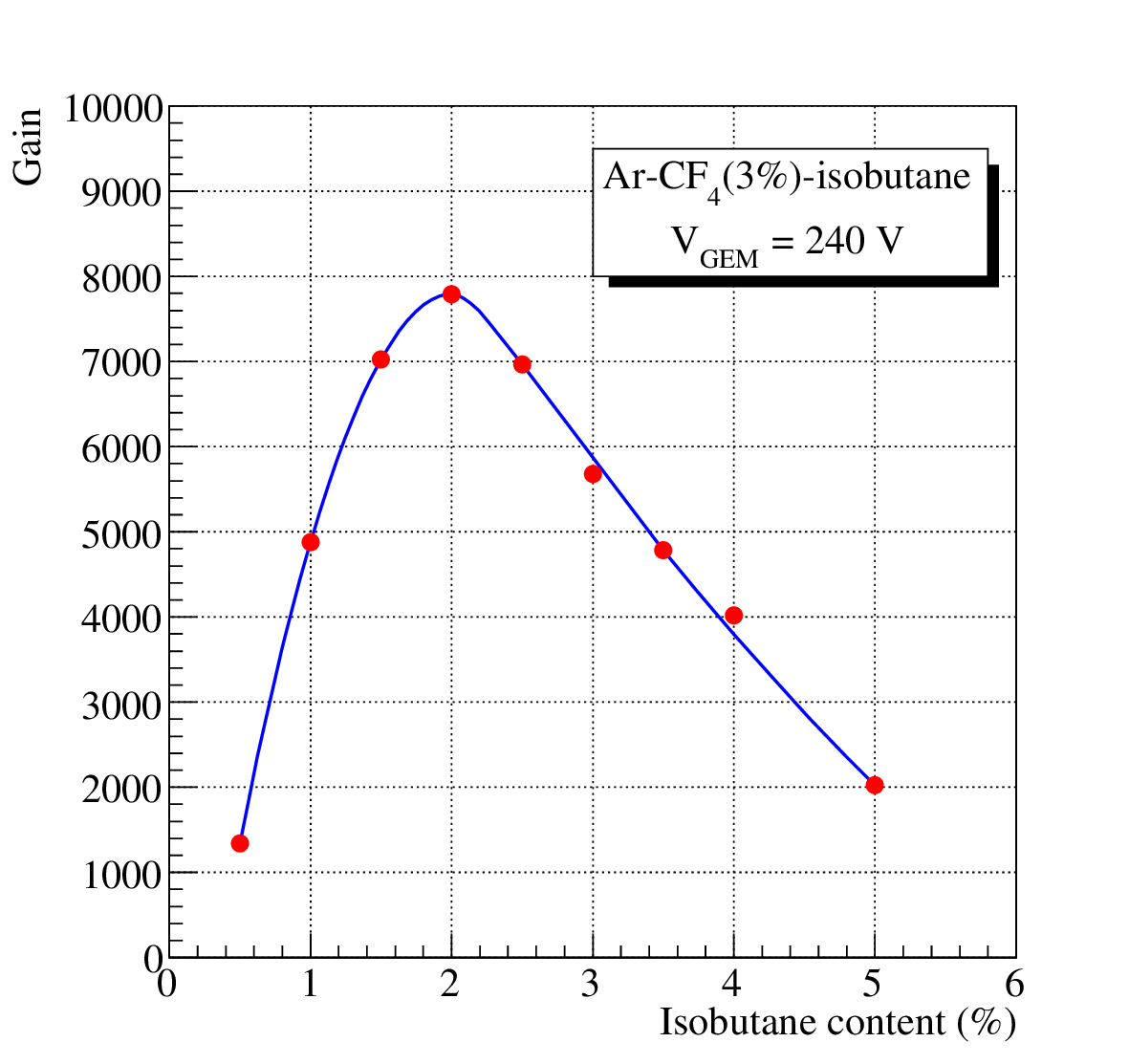}
\end{center}
\caption{\label{fig7}
\footnotesize Measured gas gain as a function of isobutane concentration.
Gas: Ar-CF$_4$(3\%)-isobutane, V$_{\rm GEM}$ = 240 V. 
The curve is only to guide the eye.
}
\end{figure}

The results are shown in Fig.~\ref{fig6} for Ar-CF$_4$(3\%) and
Ar-CF$_4$(3\%) with 2\% admixture of
methane, ethane, propane or isobutane.\footnote 
{
The CF$_4$ concentration is fixed to 3\% in the present study somewhat
arbitrarily. It should be noted, however, that a higher CF$_4$ content
requires a higher drift field corresponding to the drift-velocity maximum.
}
The gain obtained with a P10 gas (Ar-CH$_4$(10\%)) is also included as a reference.
Figure~\ref{fig7} gives the measured gas gain as a function of the isobutane content.
The signal could not be observed without isobutane at V$_{\rm GEM}$ = 240 V.

\begin{figure}[htbp]
\begin{center}
\hspace{10mm}
\includegraphics*[scale=0.57]{\figdir/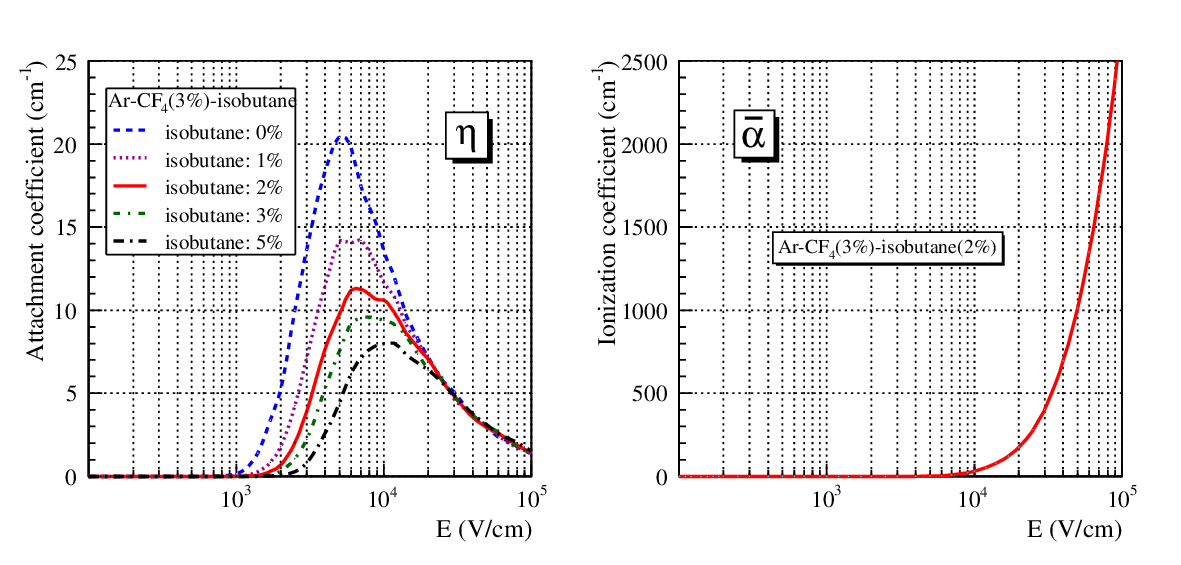}
\end{center}
\caption{\label{fig8}
\footnotesize Attachment rates (left) and effective ionization coefficient (right)
as function of the electric field given by Magboltz.    
}
\end{figure}

The gain study shows that the addition of a small amount of isobutane (or propane)
to Ar-CF$_4$ drastically lowers the operating GEM voltages which is
required to obtain an adequate gas gain,
as has already been found~\cite{CF4-papers-2}.
This is most likely due to the Penning effect since methane (ethane) is
much less (moderately) effective as an additive gas.\footnote
{
The first metastable excited state of argon is at 11.55 eV while the 
ionization potentials of methane, ethane, propane and isobutane are
12.75, 11.49, 11.07 and 10.78 eV, respectively.  
}
In addition, Magboltz shows that a small content of isobutane reduces
the electron attachment by CF$_4$ molecules in the transfer/induction
gaps by cooling down the migrating electrons, thereby increasing the
effective gain of a GEM stack (see Fig.~\ref{fig8}).
It should be noted that high electric fields in the transfer and/or induction gaps
could cause a significant loss of effective gain (even with isobutane).
In the amplification region ($E \ge 10$ kV/cm) the gas multiplication dominates.   
 
From the measurement above we decided to carry out the subsequent
performance tests with the prototype TPC using a mixture of
Ar-CF$_4$(3\%)-isobutane(2\%), 
with V$_{\rm GEM}$ = 240 V and the transfer and the induction fields of 
1.6 kV/cm, resulting in an effective gas gain of $\sim$ 8000.

\section{Performance of prototype}

The data analysis of the recorded cosmic ray events was carried out using the traditional
MultiFit (DoubleFit) program~\cite{MultiFit}.
The hit coordinate on each pad row was determined by a simple charge centroid
(barycenter) method using the pad signals within a charge cluster in the row,
without any correction to the possible bias due to the finite pad pitch. 
Our major concern is the spatial resolution along the pad-row direction
($\sigma_{\rm X}$) since the requirement to the resolution along the drift direction
($\sigma_{\rm Z}$) is moderate for the LCTPC. 
The resolution $\sigma_{\rm Z}$ was measured to be 400 $\mu$m - 700 $\mu$m,
depending on the drift distance and the drift-field strength. 

We present in this section the drift properties of the gas and the performance of
the prototype TPC obtained using cosmic rays
nearly perpendicular ($\pm\;2^\circ$) to the readout pad rows
and parallel ($\pm\;8^\circ$) to the pad plane, unless otherwise stated.

\subsection{Drift properties}

\begin{figure*}[t]
\begin{center}
\hspace{10mm}
\includegraphics*[scale=0.90]{\figdir/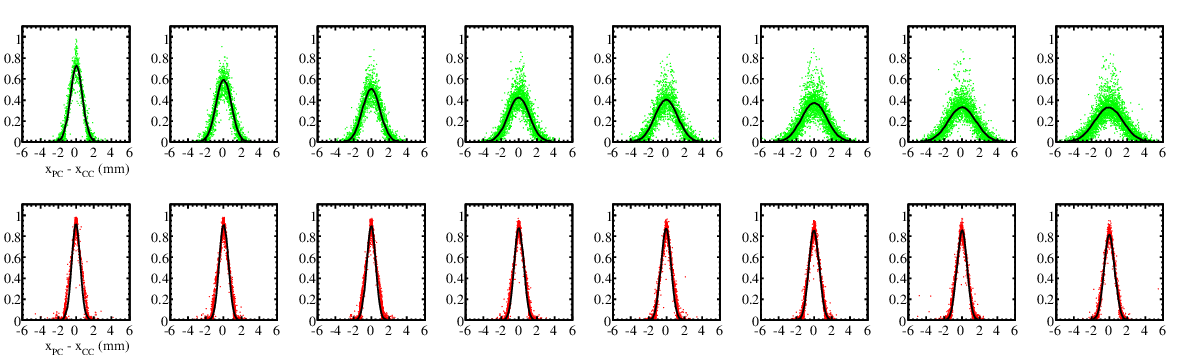}
\end{center}
\caption{\label{fig9}
\footnotesize Pad responses for different drift distances, increasing from the 
left to the right up to 257 mm.
The upper (lower) plots are for $B$ = 0 T ($B$ = 1 T).
Gas: Ar-CF$_4$(3\%)-isobutane(2\%), Drift field: 130 V/cm.
See text for the definitions of the horizontal and vertical axes.  
The solid curves represent the gaussians fitted through the data points.
}
\end{figure*}

Examples of the observed pad responses are plotted in Fig.~9,
where the normalized charge on the pad 
($q_i / \sum_j q_j$, with $q_i$ being the signal charge on pad $i\,$)
is plotted against the distance
of the pad center measured from the charge centroid, on a track by track basis.
The data points are fitted with a gaussian (solid curve in the figure) in order
to obtain the pad-response width for each region of the drift distance.     
Fig.~10 shows the pad-response width squared ($\sigma_{\rm PR}^2$) as a
function of the drift distance ($z\/$) and a fitted straight line. 
The diffusion constants ($D\/$) were determined using the relation
\begin{equation}
\sigma_{\rm PR}^2 = \sigma_{\rm PR0}^2 + D^2 \cdot z \;.
\end{equation}
The constant term is given by 
\begin{equation}
\sigma_{\rm PR0}^2 = \frac{w^2}{12} + \sigma_{\rm PRF}^2 \;,
\end{equation}
where $w$ denotes the pad pitch (1.27 mm) and $\sigma_{\rm PRF}$ is the
width of the avalanche-charge spread in the GEM stack along the pad row
direction, for single drift electrons~\cite{Paul,KEKPrep}.
The intersects of the fitted straight lines were always slightly larger
than those calculated using Eq.~(4) assuming the diffusion constants given by
Magboltz in the transfer and induction gaps of the GEM stack
to estimate $\sigma_{\rm PRF}$.
The slopes of the fitted lines give the diffusion constants, which are
summarized in Table 1 along with the measured drift velocities. 

The drift velocity was determined from the full width of the drift-time
distribution and the maximum drift length (257 mm).
Figure~11 shows an example of the drift time distribution.
The errors of the measured drift velocities were estimated conservatively 
assuming reading errors of $\pm$ 1 bin (80 nsec) both for the rising and falling
edges of the distribution. 

\begin{figure}[htbp]
\begin{center}
\hspace{10mm}
\includegraphics*[scale=0.50]{\figdir/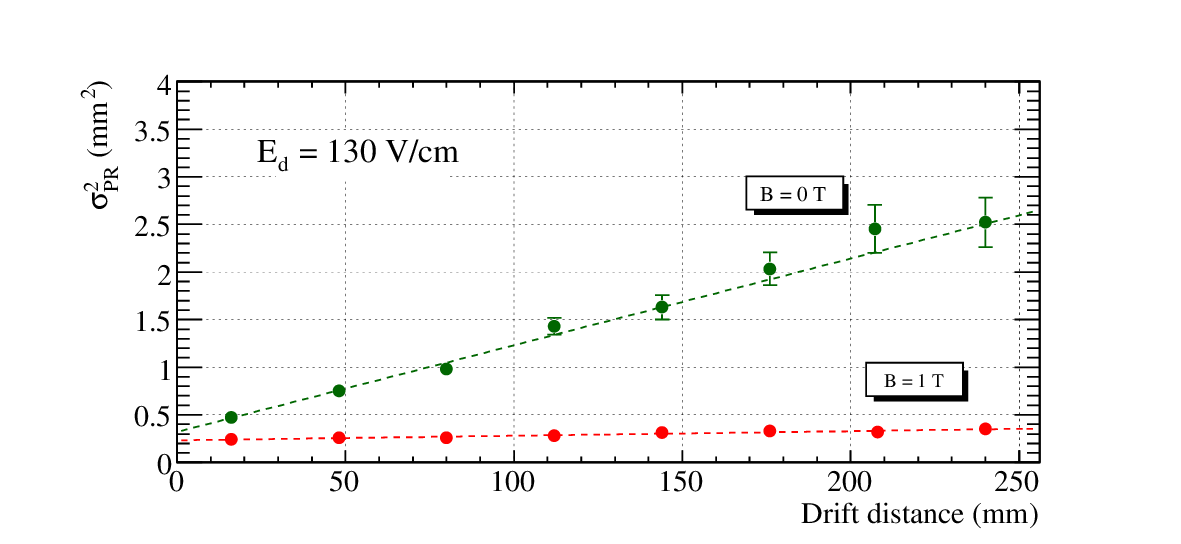}
\end{center}
\caption{\label{fig10}
\footnotesize Pad-response width squared vs. drift distance.
Gas: Ar-CF$_4$(3\%)-isobutane(2\%), Drift field: 130 V/cm. 
}
\end{figure}

\begin{figure}[htbp]
\begin{center}
\hspace{10mm}
\includegraphics*[scale=0.33]{\figdir/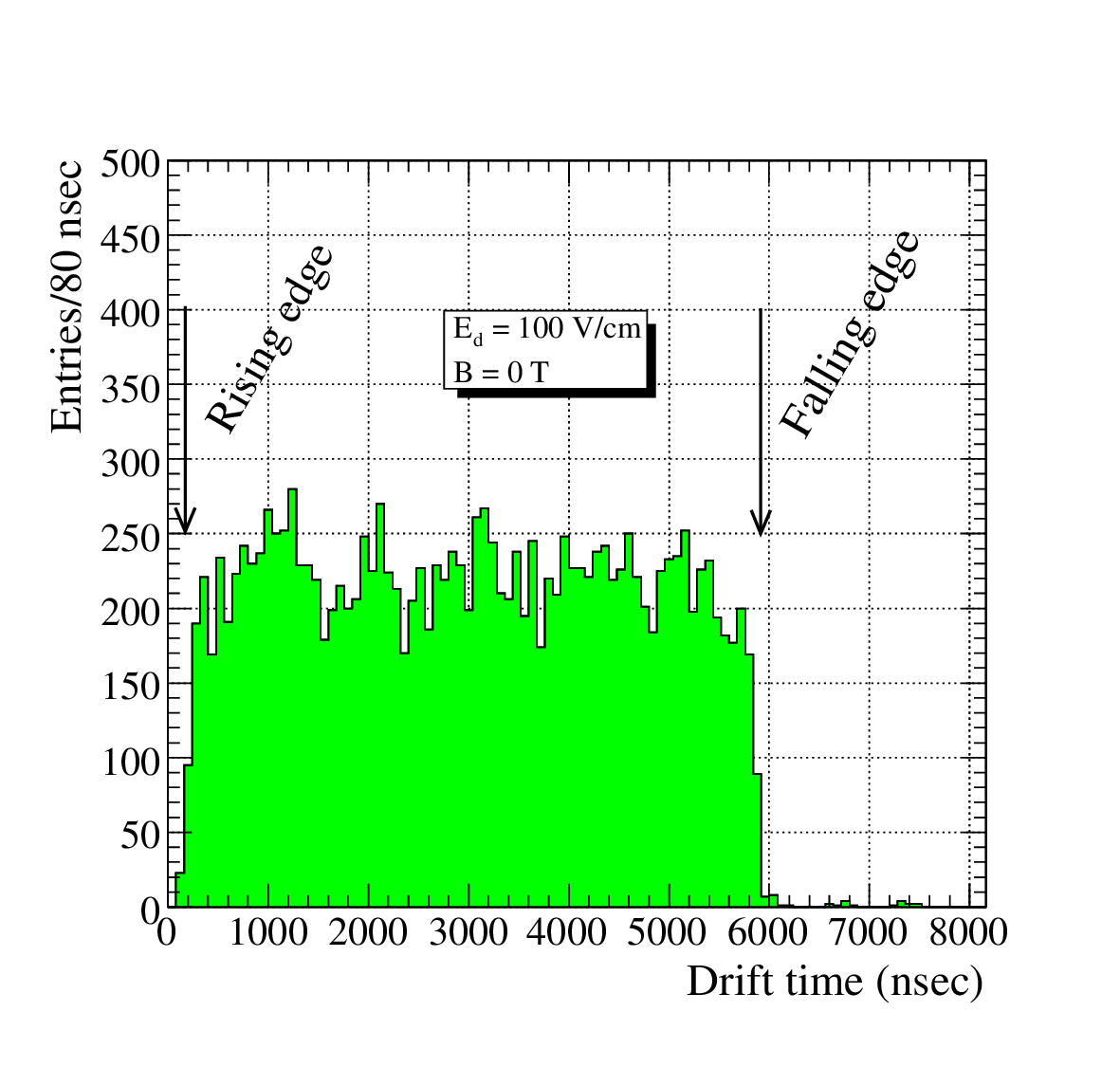}
\end{center}
\caption{\label{fig11}
\footnotesize Drift time distribution.
Gas: Ar-CF$_4$(3\%)-isobutane(2\%), Drift field: 100 V/cm, B = 0 T. 
}
\end{figure}

The measured drift properties are plotted in Fig.~12 along with the
corresponding Magboltz predictions, which closely reproduce the data points.
The reliability of Magboltz for this gas mixture was thus confirmed for the
practical range of the electric field for TPC operation, i.e. between those
corresponding to the diffusion minimum and the drift-velocity maximum.

\begin{figure}[htbp]
\begin{center}
\hspace{10mm}
\includegraphics*[scale=0.45]{\figdir/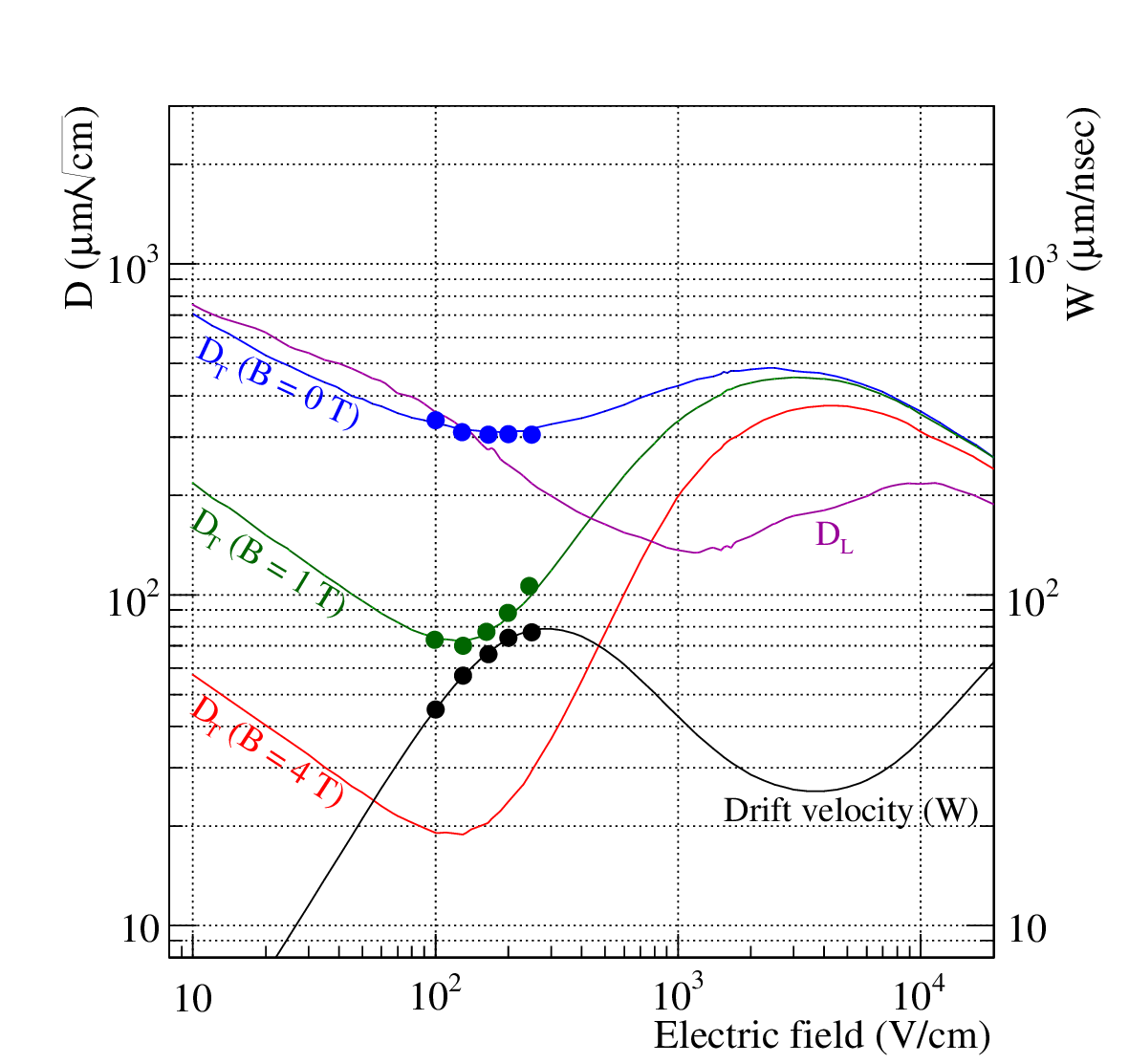}
\end{center}
\caption{\label{fig12}
\footnotesize Measured transverse diffusion constants and drift velocities.
Gas: Ar-CF$_4$(3\%)-isobutane(2\%).
The Magboltz predictions (curves) are also shown for comparison. 
}
\end{figure}

\input{Table1.tex}

\subsection{Spatial resolution}

\begin{figure}[htbp]
\begin{center}
\hspace{10mm}
\includegraphics*[scale=0.55]{\figdir/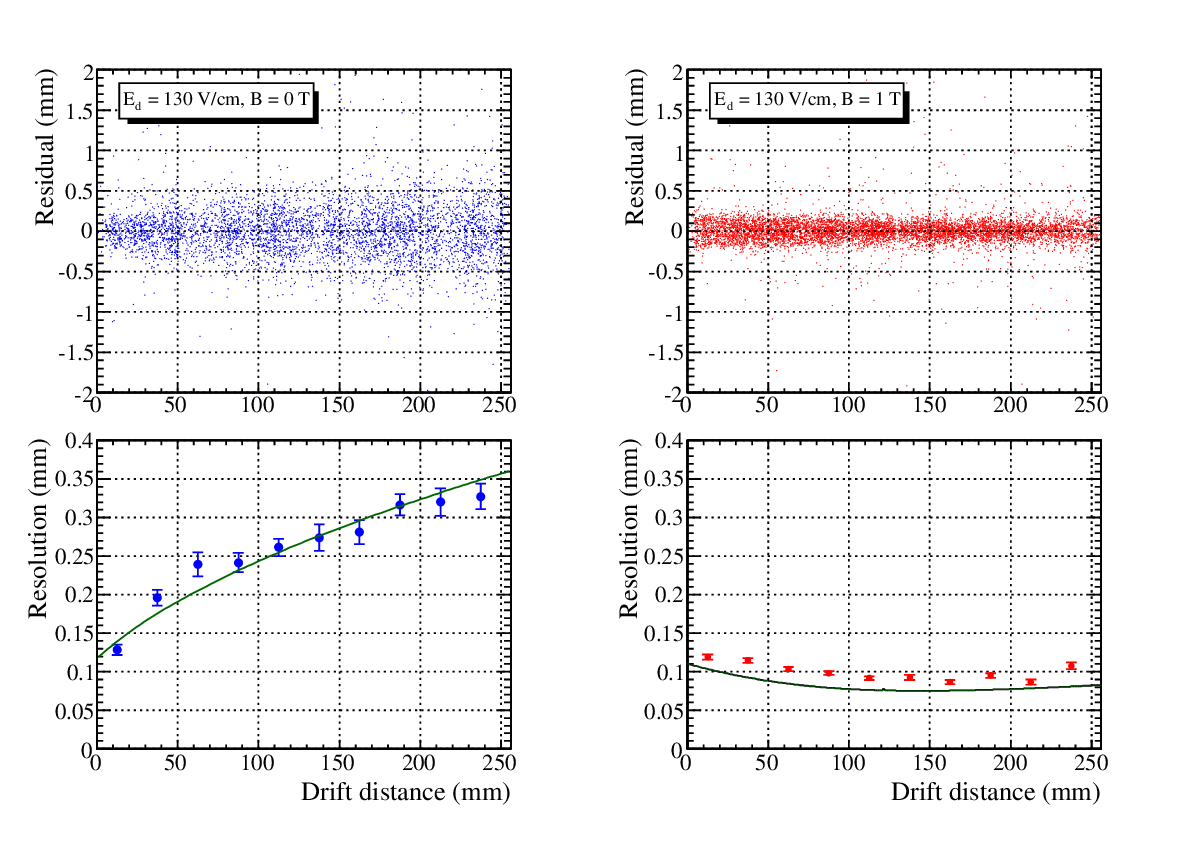}
\end{center}
\caption{\label{fig13}
\footnotesize Geometric mean residual vs. drift distance
for $B$ = 0 T (left) and $B$ = 1 T (right).
Gas: Ar-CF$_4$(3\%)-isobutane(2\%), Drift field: 130 V/cm.
The widths of the residuals are plotted against the drift distance in the lower figures,
with a  fitted curve (Eq.~(1)) for $B$ = 0 T and a calculated resolution
for $B$ = 1 T, respectively.  
}
\end{figure}

The spatial resolution along the pad row direction was obtained by plotting
the geometric mean residuals against the drift distance on an event-by-event basis.
A demonstration of the geometric mean method and the definition of the 
geometric mean residual are given in Appendix A. 
Figure~13 shows examples of the scatter plots thus obtained and the width along 
the ordinate axis as a function of the drift distance.
In order to estimate the widths (standard deviations) the scatter plot was binned by
drift distance and fitted by gaussians. 
The figures clearly show the improvement of the resolution due to a 1 T axial
magnetic field. 
It should be noted that the resolution obtained with the magnetic field can not be 
fitted with Eq.~(1) because of the degradation at small drift distances 
due to the finite pad-pitch effect\footnote{
This effect is much less prominent in the absence of magnetic field
because of larger diffusion constants.
}.
The curve shown in the figure for $B$ = 1 T is the result of the analytic 
calculation described in Appendix B and in Refs.~\cite{Paul,KEKPrep}. 

\begin{figure}[htbp]
\begin{center}
\hspace{10mm}
\includegraphics*[scale=0.45]{\figdir/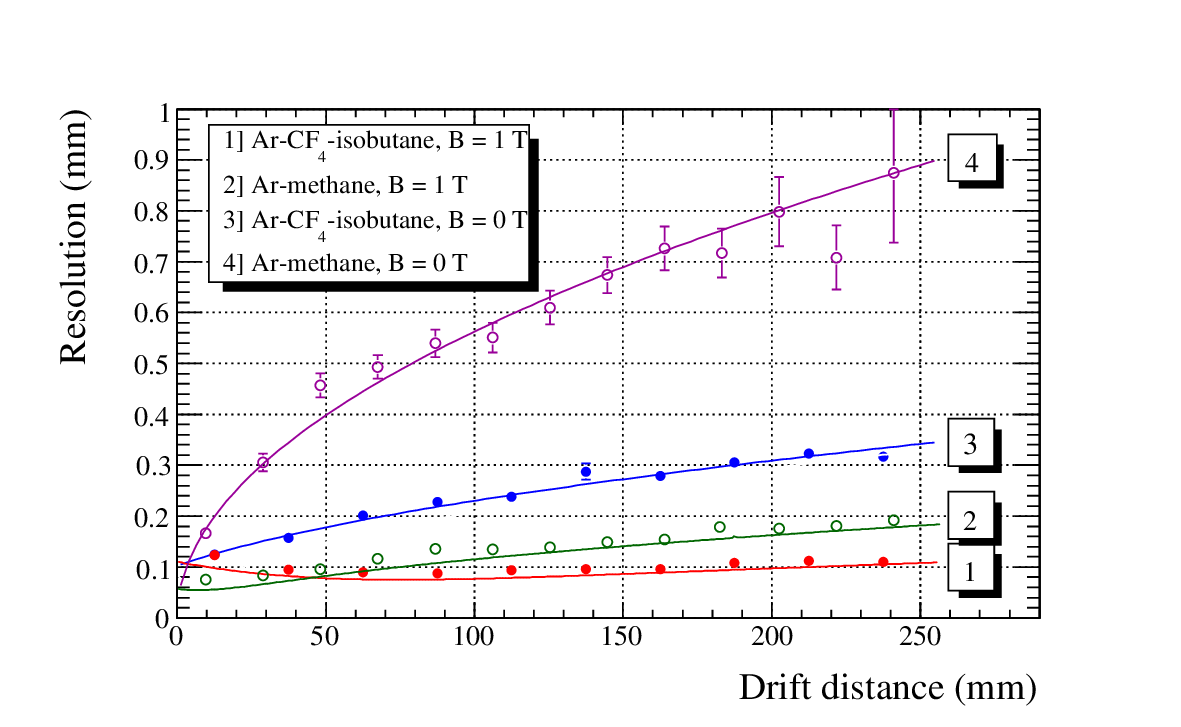}
\end{center}
\caption{\label{fig14}
\footnotesize Spatial resolution vs. drift distance.
Gas: Ar-CF$_4$(3\%)-isobutane(2\%), Drift field: 250 V/cm.
Data points for P5 gas~\cite{Kobayashi2} are also plotted for comparison.
Gas: Ar-CH$_4$(5\%), Drift field: 100 V/cm. 
The curves in the figure are Eq.~(1) fitted through the data points for $B$ = 0 T,
and calculated resolutions for $B$ = 1 T.
}
\end{figure}

In Fig.~14 the resolutions are compared to those measured previously with a
P5 gas in the beam test~\cite{Kobayashi2} for drift fields corresponding
approximately to the drift-velocity maxima.
The figure shows the smaller degradation of the resolution at long drift distances
due to diffusion in Ar-CF$_4$-isobutane.
The effective numbers of electrons ($N_{\rm eff}$) for $B$ = 0 T listed in Table 1
were obtained readily from $D^2$ given by the pad response (Eq.~(3)) and
$D^2/N_{\rm eff}$ determined from the spatial resolution (Eq.~(1)),
as function of the drift distance.
The average value of $N_{\rm eff}$ for Ar-CF$_4$(3\%)-isobutane(2\%)
is $22.6 \pm 1.5$,
which should be compared to $22 \pm 2$ obtained with a P5 gas~\cite{Kobayashi2}.
All these values are comparable to an estimation~\cite{Kobayashi}.
It should be noted, however,  that the apparent values of  $N_{\rm eff}$ could be
smaller than the {\it true} values because of electronic noise and
inadequate calibration of the readout electronics (see Appendix C).
Therefore the measured values are the lower limits of the genuine $N_{\rm eff}$
determined by the primary ionization statistics and the avalanche fluctuation.

As mentioned above it is difficult to deduce the value of $N_{\rm eff}$ using
Eq.~(1) for $B$ = 1 T because of the finite pad-pitch effect
unless the drift region is long enough.
Actually the observed resolution is the quadratic sum of the diffusion contribution, 
the finite pad-pitch term and a constant offset ($\sim \sigma_{\rm X0}$)\footnote{
Strictly speaking, the constant term ($\sigma_{\rm X0}$) is slightly greater than the
constant offset since the diffusion term contains small but finite contribution to the
constant term (see Eq.~(\ref{b5}) in Appendix B). 
}.
The last term, which may depend on experimental conditions,
can be estimated as the difference of the measured resolution squared
and the calculated one.

As an example, each component contributing to the resolution is shown in Fig.~15
along with the measured resolution for $E_{\rm d}$ = 250 V/cm.
In the calculation the diffusion constants in the drift region and in the
transfer/induction gaps given by Magboltz were used.
The value of $N_{\rm eff}$ was assumed to be 23.
The constant offset is about (48 $\mu$m)$^2$ and is due most likely to
electronic noise and inadequate calibration of the readout electronics.
Possible major sources of the constant term are briefly discussed in Appendix C.

\begin{figure}[htbp]
\begin{center}
\hspace{10mm}
\includegraphics*[scale=0.50]{\figdir/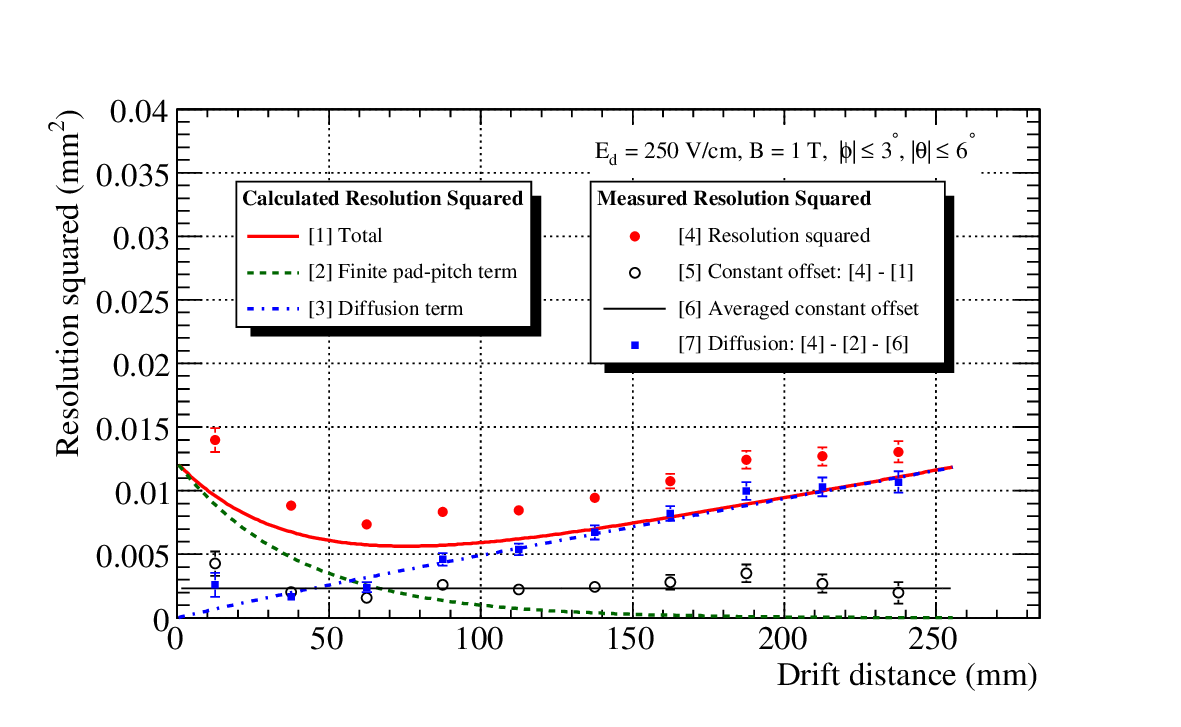}
\end{center}
\caption{\label{fig15}
\footnotesize Spatial resolution squared vs. drift distance.
Gas: Ar-CF$_4$(3\%)-isobutane(2\%), Drift field: 250 V/cm, $B$ = 1 T.
The curves are the calculated resolution squared (solid line) and its
components: the finite pad-pitch term (dashed line) and
the diffusion term (dash-dotted line). 
The horizontal straight line represents the fitted constant offset.
}
\end{figure}

The net diffusion contribution can thus be extracted by subtractions
and is shown in filled squares in the figure.
By fitting a straight line through those data points,
$N_{\rm eff}$ under the 1 T magnetic field could be estimated.
The resultant values are about 21 and 23, depending on the
diffusion constant assumed, the Magboltz value (100.9 $\mu$m/$\sqrt{\rm cm}$) or
the measured value (106 $\mu$m/$\sqrt{\rm cm}$)\footnote{
The diffusion constant and the value of $N_{\rm eff}$ assumed in the calculation
do not affect the resultant $N_{\rm eff}$ since they are used only to estimate
the {\it constant\/} offset.
}.
The values of $N_{\rm eff}$ thus obtained is consistent with those found
from the resolutions in the absence of magnetic field (see Table~1),
indicating that the effective number of electrons is not affected by 
the existence of the magnetic field as well as by the variation of the 
drift field.

\section{Expected resolution of an LCTPC}

Our measurements of the drift properties in Ar-CF$_4$(3\%)-isobutane(2\%) 
are well reproduced by the Magboltz code for $B$ = 0 and 1 T.
This indicates that the cross sections of the gas molecules involved are 
correctly incorporated in the code at least for the corresponding electron energies
for Ar-CF$_4$-isobutane seen in Fig.~3
and that the influence of the magnetic field on the transverse 
diffusion is properly included.
Since the energy (random velocity) distribution of drifting electrons is not
affected by the existence of axial magnetic field Magboltz is expected to give
correct values of diffusion constants under stronger magnetic fields.
In addition, the effective number of electrons is measured to be about 23 for the
pad row pitch of 6.3 mm, which is affected little by the magnetic field if any.
Therefore it is now possible to estimate the spatial resolution of the LCTPC
using Eq.~(\ref{b4}) in Appendix B.

Figure~16 shows the expected resolutions as function of the drift distance up to
2.5 m for a GEM-equipped LCTPC. 
In the calculation of the resolutions the transfer gaps and the induction gap
of 4 mm are assumed for a triple GEM stack, with the electric fields in the gaps of
1600 V/cm while the drift field is set to 250 V/cm corresponding approximately to
the drift-velocity maximum.
The resolution better than 100 $\mu$m is thus feasible throughout the sensitive volume
for tracks perpendicular to the pad row under an axial magnetic field of 4 T,
at which the diffusion constant is predicted to be 29 $\mu$m/$\sqrt{\rm cm}$ by Magboltz, 
if the constant offset is kept reasonably small.

\begin{figure}[htbp]
\begin{center}
\hspace{10mm}
\includegraphics*[scale=0.60]{\figdir/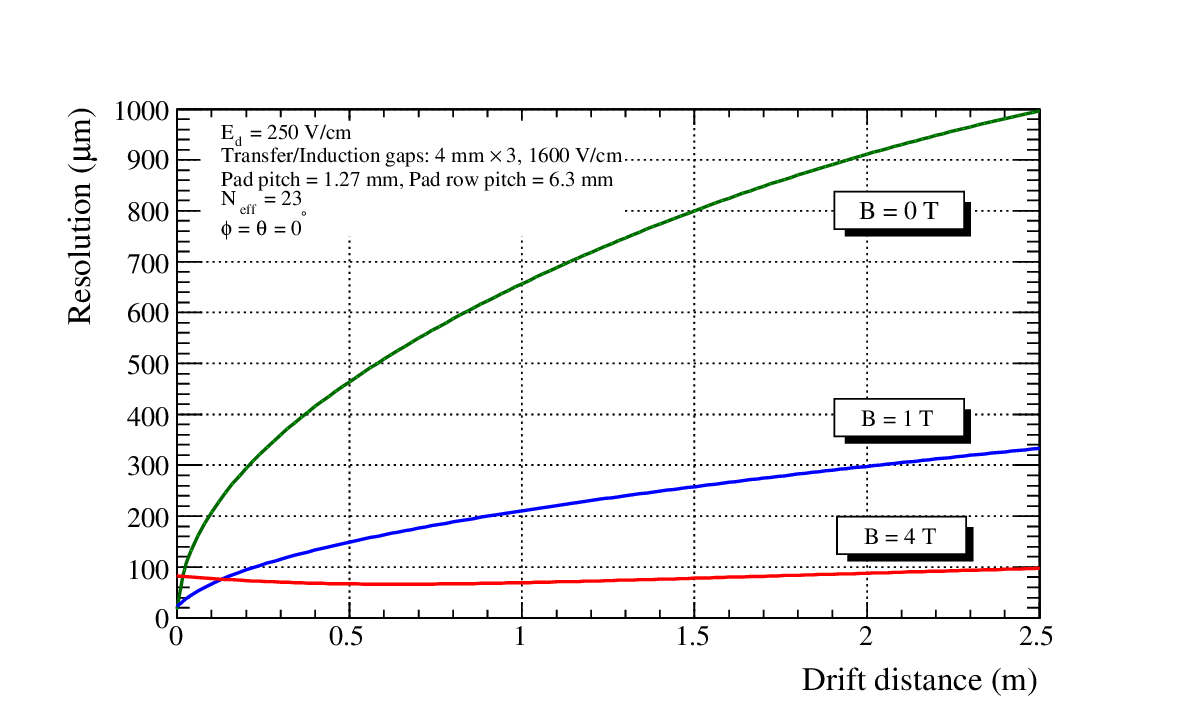}
\end{center}
\caption{\label{fig16}
\footnotesize Expected spatial resolution of a GEM-equipped LCTPC
for tracks perpendicular to the pad rows.
Gas: Ar-CF$_4$(3\%)-isobutane(2\%), Drift field: 250 V/cm, $B$ = 4 T.
The transfer and induction gaps are assumed to be 4 mm.
}
\end{figure}

\section{Conclusions}

We have measured the diffusion constant, the drift velocity and 
the spatial resolution, as well as the gas gain, using a GEM-equipped
TPC prototype operated in a gas mixture of Ar-CF$_4$(3\%)-isobutane(2\%)
under magnetic fields of 0 T and 1 T.
The prototype has been working stably with a sufficient gas gain in the
cosmic ray test
and gave a reasonable effective number of electrons ($N_{\rm eff}$),
indicating a small loss, if any, of drift electrons at the entrance
to the GEM stack.
A reasonable value of $N_{\rm eff}$ means also that the gas-gain (avalanche)
fluctuation is moderate.
A small amount of isobutane ($\sim$ 2\%) was found to be very effective
to lower the operating high voltages of the GEMs, most likely due to the
Penning effect. 

The measured diffusion constants ($D$) in Ar-CF$_4$(3\%)-isobutane(2\%)
are consistent with those given
by Magboltz both for $B$ = 0 T and $B$ = 1 T, demonstrating the
reliability of the Magboltz predictions for gas mixtures containing
argon, CF$_4$ and isobutane.
This gas mixture is confirmed to be cooler and faster than conventional
TPC gases such as P5 and to ensure better azimuthal spatial resolution
at long drift distances under a high axial magnetic field. 

The spatial resolution better than 100 $\mu$m, the target of the LCTPC,
is expected to be within reach with a GEM-based TPC operated in
Ar-CF$_4$(3\%) plus a small amount of isobutane under a 4 T magnetic field.

\setcounter{figure}{0}

\appendix
\section{Geometric mean method}
\input{AppendixA.tex}

\section{Analytic expression of the spatial resolution}
\input{AppendixB.tex}

\section{Contributors to the constant term}
\input{AppendixC.tex}

\section*{\nonumber Acknowledgments}
We would like to thank the group at the KEK cryogenics science center for
the preparation and the operation of the superconducting magnet.
We are also grateful to many colleagues of the LCTPC collaboration for their
continuous encouragement and support, and for fruitful discussions.
Finally, we would like to thank Prof. Stephen Biagi, 
who permitted us to reproduce his plots in Fig. 1.
This work was supported by the Creative Scientific Research
Grant No. 18GS0202 of the Japan Society of Promotion of Science.

\newpage

\input{addendum.tex}

%
\end{document}

%% file: authorlist.tex
%
%
\author[3]{M.~Kobayashi\corref{cor1}}
     \ead{makoto.kobayashi.exp@kek.jp}
     \cortext[cor1]{Corresponding author.
                           Tel.: +81 29 864 5379; fax: +81 29 864 2580.}
\author[7]{R.~Yonamine}
\author[2]{T.~Tomioka}
\author[1]{A.~Aoza}
\author[2]{H.~Bito}
\author[3]{K.~Fujii}
\author[1]{T.~Higashi}
\author[4]{K.~Hiramatsu}
\author[3]{K.~Ikematsu}
\author[1]{A.~Ishikawa}
\author[4]{Y.~Kato}
\author[1]{H.~Kuroiwa}
\author[3]{T.~Matsuda}
\author[2]{O.~Nitoh}
\author[2]{H.~Ohta}
\author[2]{K.~Sakai}
\author[5]{R.D.~Settles}
\author[1]{A.~Sugiyama}
\author[1]{H.~Tsuji}
\author[6]{T.~Watanabe}
\author[3]{H.~Yamaoka}
\author[4]{T.~Yazu}
%
%
\address[3]{High Energy Accelerator Research Organization (KEK), Tsukuba, 305-0801, Japan}
\address[7]{Graduate University for Advanced Studies, KEK, Tsukuba, 305-0801, Japan}
\address[2]{Tokyo University of Agriculture and Technology, Koganei, 184-8588, Japan}
\address[1]{Saga University, Saga, 840-8502, Japan}
\address[4]{Kinki University, Higashi-Osaka, 577-8502, Japan}
\address[5]{Max Planck Institute for Physics, DE-80805 Munich, Germany}
\address[6]{Kogakuin University, Hachioji, 192-0015, Japan}

%% file: Table1.tex
\begin{table}[htbp]
\begin{center}

\caption{\label{tb1:table1}
 Drift properties and $N_{\rm eff}$ in Ar-CF$_4$(3\%)-isobutane(2\%)}

\bigskip

{\footnotesize (a) $B$ = 0 T \\}
\medskip
\begin{tabular}{|c||c|c||c|c||c||c|} \hline
 &  \multicolumn{2}{|c||}{Drift velocity (cm/$\mu$s)} & 
 \multicolumn{2}{|c||}{Diffusion constant ($\mu$m/$\sqrt{{\rm cm}}$)} & $D/\sqrt{N_{{\rm eff}}}$& \\
\cline{2-5}
$E$ (V/cm)& Measured & Magboltz & Measured & Magboltz & ($\mu$m/$\sqrt{{\rm cm}}$) & $N_{\rm eff}$\\
\hline \hline
100 & 4.45 $\pm$ 0.13 & 4.50 & 337 $\pm$ 12 & 327.8 $\pm$ 2.4 & 74 $\pm$ 2 & 20.7 $\pm$ 1.8 \\ \hline
130 & 5.67 $\pm$ 0.20 & 5.67 & 310 $\pm$ 11 & 316.7 $\pm$ 1.9 & 67 $\pm$ 2 & 21.1 $\pm$ 1.9 \\ \hline
165 & 6.61 $\pm$ 0.28 & 6.69 & 305 $\pm$ \hspace{0.40mm} 8 & 311.7 $\pm$ 2.5 & 62 $\pm$ 2 & 24.1 $\pm$ 1.5 \\ \hline
200 & 7.38 $\pm$ 0.34 & 7.35 & 306 $\pm$ \hspace{0.40mm} 8 & 311.2 $\pm$ 2.2 & 63 $\pm$ 2 & 23.8 $\pm$ 1.4 \\ \hline
250 & 7.74 $\pm$ 0.38 & 7.81 & 305 $\pm$ 10 & 315.4 $\pm$ 2.4 & 65 $\pm$ 2 & 22.0 $\pm$ 1.8\\
\hline

\end{tabular} 


\bigskip 

{\footnotesize (b) $B$ = 1 T \\}
\medskip
\begin{tabular}{|c||c|c||c|c|} \hline
 &  \multicolumn{2}{|c||}{Drift velocity (cm/$\mu$s)} & 
 \multicolumn{2}{|c|}{Diffusion constant ($\mu$m/$\sqrt{{\rm cm}}$)} \\
\cline{2-5}
$E$ (V/cm)& Measured & Magboltz & Measured & Magboltz \\
\hline \hline
100 & 4.52 $\pm$ 0.13 & 4.50 & \hspace{2.4mm} 72 $\pm$ 2 & \hspace{0.9mm} 73.8 $\pm$ 0.7 \\ \hline
130 & 5.74 $\pm$ 0.20 & 5.67 & \hspace{2.4mm} 70 $\pm$ 2 & \hspace{0.9mm} 72.8 $\pm$ 0.6 \\ \hline
165 & 6.69 $\pm$ 0.28 & 6.69 & \hspace{2.4mm} 77 $\pm$ 2 & \hspace{0.9mm} 77.3 $\pm$ 0.6 \\ \hline
200 & 7.30 $\pm$ 0.33 & 7.35 & \hspace{2.4mm} 88 $\pm$ 2 & \hspace{0.9mm} 85.5 $\pm$ 0.7 \\ \hline
250 & 7.65 $\pm$ 0.36 & 7.81 & \hspace{0.70mm} 106 $\pm$ 2 & 100.9 $\pm$ 0.7 \\
\hline

\end{tabular} 


\end{center}

\end{table}

%% file: AppendixA.tex
In this appendix we give a simple demonstration of the geometric mean method
applied in the analysis to estimate the spatial resolution.

First, in the case where the hit point in question 
(say, $x_i$, on the $i$-th pad row) 
is {\it excluded\/} in the track fitting,
the residual is given by 
\begin{displaymath}
 \Delta x_i = x_i - \hat{x}_i\;,
\end{displaymath}
where $\hat{x}_i$ represents the estimator for the $i$-th point
given by the track fitting using the remaining hit points.
Since $x_i$ and $\hat{x}_i$ are statistically independent
the variance of the residual is given by
\begin{eqnarray}
\sigma_{\rm excl}^2 &\equiv& \left<(x_i - \hat{x}_i)^2\right>
                                - \left<x_i - \hat{x}_i \right>^2 \nonumber \\        
      &=& \left< x_i^2 - 2x_i\hat{x}_i + \hat{x}_i^2 \right>
                              - \left< x_i - \hat{x}_i \right>^2   \nonumber \\
      &=& \left<x_i^2\right> - \left<x_i \right>^2
                 + \left<\hat{x}_i^2\right> - \left<\hat{x}_i \right>^2 \nonumber \\
      &=&        \sigma_{x_i}^2 + \sigma_{\hat{x}_i}^2 \label{a1}\;,
\end{eqnarray}   
the sum of the {\it true\/} spatial resolution and the tracking error.

Next, in the case where the hit point in question is {\it included\/}
in the track fitting,
the estimator for the $i$-th hit point is the weighted mean of
$\hat{x}_i$ and $x_i$:
\begin{displaymath}
 \hat{x}_i^\prime = \frac{w_{\hat{x}_i} \hat{x}_i + w_{x_i} x_i}
                        {w_{\hat{x}_i} +  w_{x_i}} \;,
\end{displaymath}   
with $w_{\hat{x}_i}$ $(w_{x_i})$ being the corresponding weight:
$1/{\sigma_{\hat{x}_i}^2}$ $(1/{\sigma_{x_i}^2})$.
The residual is hence given by 
\begin{eqnarray}
 \Delta x_i^\prime \equiv x_i - \hat{x}_i^\prime &=& 
    \frac{\sigma_{x_i}^2}{\sigma_{x_i}^2 + \sigma_{\hat{x}_i}^2}
     (x_i - \hat{x}_i) \nonumber \\
 &=& \frac{\sigma_{x_i}^2}{\sigma_{x_i}^2 + \sigma_{\hat{x}_i}^2}
         \cdot \Delta x_i \label{a2}\;.
\end{eqnarray}
The variance of the residual in this case is therefore
\begin{eqnarray}
\sigma_{\rm incl}^2 &\equiv&
                 \left< (x_i - \hat{x}_i^\prime)^2 \right>
                              - \left<x_i - \hat{x}_i^\prime\right>^2 \nonumber \\
  &=& \left( \frac{\sigma_{x_i}^2}{\sigma_{x_i}^2 + \sigma_{\hat{x}_i}^2} \right)^2
      \cdot 
            \left\{ \left<(x_i - \hat{x}_i)^2\right>
                                - \left<x_i - \hat{x}_i \right>^2 \right\} \nonumber \\
  &=& \left( \frac{\sigma_{x_i}^2}{\sigma_{x_i}^2 + \sigma_{\hat{x}_i}^2} \right)^2
      \cdot \sigma_{\rm excl}^2 \label{a3}\;.
\end{eqnarray}
Combining Eq.~(\ref{a1}) and Eq.~(\ref{a3}), 
the {\it true} spatial resolution is given by
\begin{equation}
\sigma_{x_i} = \sqrt{\sigma_{\rm excl} \cdot \sigma_{\rm incl}} \label{a4}\,.
\end{equation}

Equation~(\ref{a4}) demonstrates that the spatial resolution is obtained by 
taking the geometric mean of the two widths: one for the $\Delta x_i$ distribution
and the other for the $\Delta x_i^\prime$ distribution.
One may instead plot, {\it on an event-by-event basis\/}, geometric mean residuals defined as
$\Delta x_i^{\prime\prime} \equiv 
    S \cdot \sqrt{(x_i - \hat{x}_i) \cdot (x_i - \hat{x}_i^\prime)}\;,$
with $S$ being the sign of $x_i - \hat{x}_i$
(or equivalently $x_i - \hat{x}_i^\prime$\, from Eq.~(\ref{a2})),
and interpret its width as the spatial resolution since
\begin{eqnarray*}
\sigma_{\Delta x_i^{\prime\prime}}^2 &\equiv&
  \left< S^2 \cdot (x_i - \hat{x}_i) \cdot (x_i - \hat{x}_i^\prime) \right> \\
 && \hspace{10mm}
 - \left< S \cdot \sqrt{(x_i - \hat{x}_i) \cdot (x_i - \hat{x}_i^\prime)} \right>^2 \\
  &=&  \frac{\sigma_{x_i}^2}{\sigma_{x_i}^2 + \sigma_{\hat{x}_i}^2}
      \cdot \left\{ \left<(x_i - \hat{x}_i)^2\right>
                                - \left<x_i - \hat{x}_i \right>^2\right\} \\
  &=& \sigma_{x_i}^2 \;.
\end{eqnarray*}

%% file: AppendixB.tex
One way to estimate the spatial resolution of a TPC along the pad-row direction
is to write a realistic Monte-Carlo simulation code.
This technique is applicable to any complicated situation, and has been developed by several
groups.
On the other hand, an analytic approach is applicable only to a restricted
case where incident particles are normal to the pad row.
However, the resultant formula is rather simple and is sometimes
enlightening as shown below.
Though a numerical calculation is needed to evaluate the formula, the demanded
CPU time is much less than a Monte-Carlo simulation.
In addition, the analytic calculation can be used to check the reliability of a
Monte-Carlo simulation program, which is usually long and complicated.
This appendix is devoted to briefly summarize our analytic approach,
based on the following assumptions.
\begin{enumerate}
\item Particle tracks are normal to the pad row.
\item Track coordinate is determined by the charge centroid (barycenter) method.
\item Displacement of arriving drift electrons due to $E \times B$ effect
      near the entrance to the detection device is negligible.
\item Displacement of arriving electrons due to the finite granularity of
      amplification elements of the detection device
      (line intervals in MicroMEGAS or a hole pitch in GEM)
      is negligible.
\end{enumerate}

\subsection{Electronic noise contribution}

Let us first consider the electronic noise contribution to the resolution degradation.
The charge centroid of the signal including the noise is given by
\begin{displaymath}
X = \frac{1}{\sum_j (q_j + q_j^{\prime})} \cdot \sum_j (q_j + q_j^\prime) x_j^* \;,
\end{displaymath}
where $q_j$ ($q_j^{\prime}$ ) is the signal (noise) charge on pad $j$ and $x_j^*$ is 
the coordinate of the center of pad $j$.
For simplicity we assume that $\sum_j q_j^\prime \ll \sum_j q_j$, therefore
 $\sum_j (q_j + q_j^\prime) \approx \sum_j q_j$,
and that $\bigl< q_j^\prime \bigr> = 0$, 
$\bigl< q_j^\prime \cdot q_k^\prime \bigr> = \bigl< q^{\prime2} \bigr> \cdot \delta_{jk}$
and $\bigl< q_j \cdot q_k^\prime \bigr> = 0$.
Then, with $\tilde{x}$ being the true track coordinate, the squared resolution along the
pad-row direction is given by
\begin{eqnarray}
\sigma_X^2 &=& \bigl< (X-\tilde{x})^2 \bigr > \nonumber \\
&=& \biggl< \biggl(\frac{1}{\sum_j (q_j + q_j^\prime)} \cdot 
           \sum_j (q_j + q_j^{\prime}) (x_j^* - \tilde{x}) \biggr)^2 \biggr> \nonumber \\
&=& \biggl< \biggl( \frac{1}{\sum_j (q_j + q_j^\prime)} \biggr)^2 \cdot
     \biggl( \sum_j (q_j + q_j^\prime)^2( x_j^* - \tilde{x})^2  
            + \sum_{j \not\neq k} (q_j + q_j^\prime)(q_k + q_k^\prime)
                (x_j^* - \tilde{x})(x_k^* - \tilde{x})
               \biggr) \biggr> \nonumber \\
&\approx& \biggl< \biggl( \frac{1}{\sum_j q_j} \biggr)^2 \cdot
     \biggl( \sum_j (q_j^2 + q_j^{\prime2}) (x_j^* - \tilde{x})^2  
 + \sum_{j \not\neq k} q_j q_k (x_j^* - \tilde{x})(x_k^* - \tilde{x}) \biggr) \biggr> \nonumber \\
&\approx& \biggl< \biggl(\frac{1}{\sum_j q_j} \cdot 
                     \sum_j q_j (x_j^* - \tilde{x}) \biggr)^2 \biggr> 
 + \biggl< \biggl( \frac{1}{\sum_j q_j} \biggr)^2 \cdot
                              \sum_j q_j^{\prime2} (x_j^* - \tilde{x})^2 \biggr> \nonumber \\
&\approx& \biggl< \biggl(\frac{1}{\sum_j q_j} \cdot 
                     \sum_j q_j x_j^* - \tilde{x}\biggr)^2 \biggr>
 + \biggl< \frac{1}{\bigl(\sum_j q_j \bigr)^2} \biggr> \cdot
            \bigl< q^{\prime2} \bigr> \cdot \sum_j (x_j^* - \tilde{x})^2 \label{b1} \;. 
\end{eqnarray}
The second term represents the noise contribution whereas the first term gives the resolution
obtained in the ideal case in which the electronic noise is absent.

Let us define $Q_i$ as the total gas-amplified signal charge on the pad row
created by drift electron $i$.
The total charge $Q_i$ is given by $\sum_j q_{ji}$, where $q_{ji}$ is the signal charge
contribution of drift electron $i$, on pad $j$ ($q_j \equiv \sum_i q_{ji}$). 
Then $\sum_j q_j = \sum_j \sum_i q_{ji} = \sum_i Q_i$.
Assuming that the total number of drift electrons per pad row ($N$, not a constant)
is large enough
\begin{displaymath}
\sum_j q_j = \sum_{i=1}^N Q_i \approx N \cdot \bigl< Q \bigr > \;,
\end{displaymath}
where $\bigl< Q \bigr >$ is the average total signal charge per drift electron.

Now the noise contribution is given by
\begin{equation}
\sigma_X^2)_{\rm noise}
 \approx \biggl< \frac{1}{N^2} \biggr > \cdot \frac{\sigma_{q^\prime}^2}{\bigl< Q \bigr >^2}
             \cdot \sum_j (x_j^* - \tilde{x})^2 \label{b2} \;. 
\end{equation}
This term increases with the signal charge spread along the pad row direction
given by Eq.~(3)
because of the increase of the active number of pads
(the summation range of $j$ around $\tilde{x}$).
It should be noted that $x_j^*$ can be written as $j \cdot w$ with $w$ being the pad pitch. 

\subsection{Analytic evaluation of spatial resolution}

\begin{figure}[htbp]
\begin{center}
\includegraphics*[scale=0.45]{\figdir/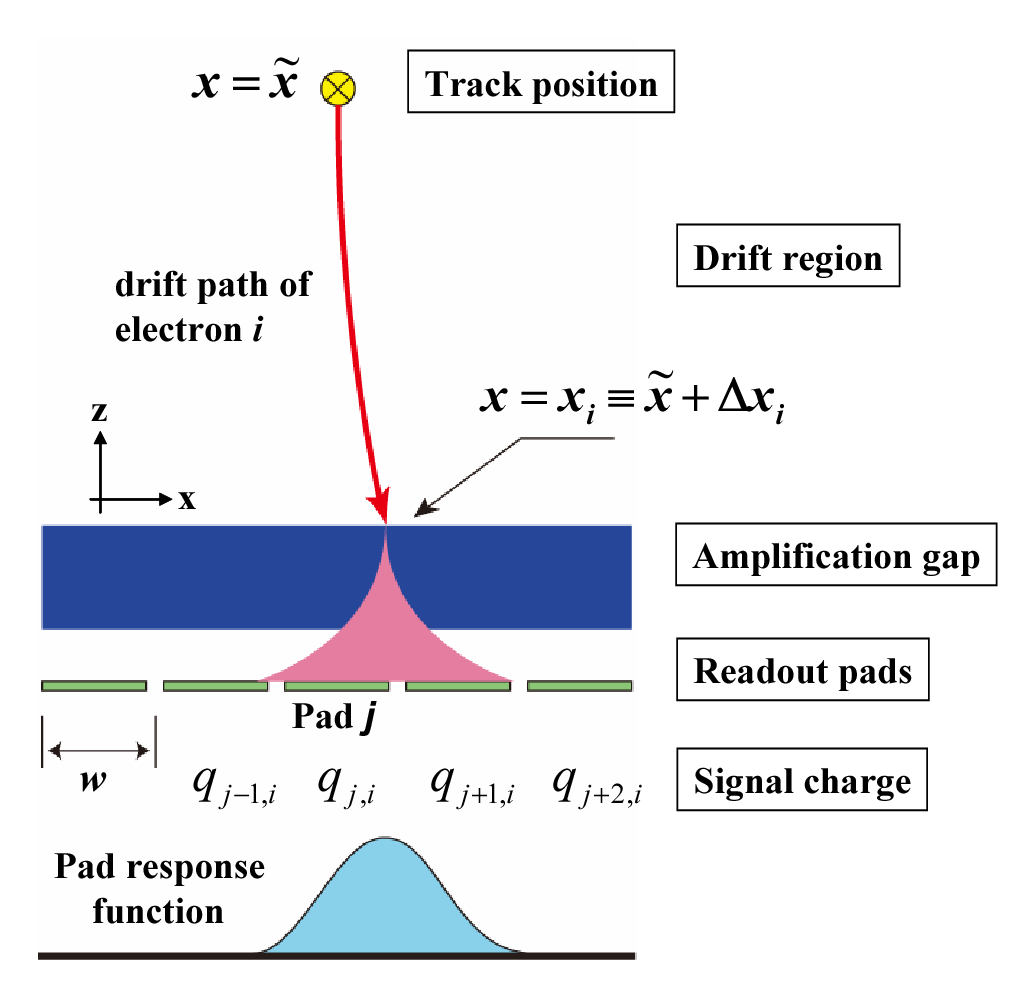}
\end{center}
\caption[figA2]{\label{figA2}
\footnotesize Illustration of the relevant variables.
}
\end{figure}

In this section we evaluate the first term in Eq.~(\ref{b1}).
Let us redefine the charge centroid as follows.
\begin{displaymath}
X = \frac{\sum_j \sum_{i=1}^N q_{ji} \cdot x_j^*}
                        {\sum_j \sum_{i=1}^{N} q_{ji}}
\equiv \frac{\sum_{i=1}^N Q_i \sum_j F_j(x_i) \cdot x_j^*}{\sum_{i=1}^N Q_i}\;,
\end{displaymath}
where (see Fig.~\ref{figA2})
\begin{eqnarray*}
N &:& {\rm total ~number ~of ~drift ~electrons ~per ~pad ~row} , \\
x_i &:& {\rm arrival~ position~ of~ electron}~ i~ {\rm at~ the~ entrance~ to 
                                              ~the~ detection~ device} , \\  
x_j^* &\equiv& j \cdot w {\rm ~~:~central~ coordinate~ of~ pad~} j \;\;\;
      (j = \cdot\cdot\cdot, -2,-1,0,+1,+2, \cdot\cdot\cdot) , \\
F_j(x_i) &\equiv& \frac{q_{ji}}{Q_i}
             \equiv \int_{jw-w/2}^{jw+w/2} f(\xi-x_i) d\xi , \\
f(\xi) &:& {\rm (normalized)~pad~ response~ function~ (PRF)}. 
\end{eqnarray*}
It should be noted that the PRF here is determined solely by the charge spread in the
detection gap and does not depend on the pad pitch whereas conventional PRFs do.

Then\footnote{
In what follows, for an arbitrary function $I$,
\begin{eqnarray*}
\bigl< I(x,Q) \bigr> &\equiv&
   \biggl( \prod_{i=1}^N \int P_x(x_i)\, dx_i \int P_Q(Q_i)\, dQ_i \biggr) I(x,Q) \;,\\
\bigl< I(x) \bigr> &\equiv&
   \bigl< I(x) \bigr>_x \equiv \biggl( \prod_{i=1}^N \int P_x(x_i)\, dx_i \biggr) I(x) \;,\\
\bigl< I(Q) \bigr> &\equiv&
   \bigl< I(Q) \bigr>_Q \equiv \biggl( \prod_{i=1}^N \int P_Q(Q_i)\, dQ_i \biggr) I(Q) \;, \\
\end{eqnarray*} 
where $P_x(x_i)$ and $P_Q(Q_i)$ denote the probability density functions,
respectively for the arrival position and the signal charge of drift electron $i$
whereas $N$ is the total number of drift electrons per pad row (not a constant, actually).
It should be noted that 
$\left< I_1(x) \cdot I_2(Q) \right > = \left< I_1(x) \right>_x \cdot \left< I_2(Q) \right>_Q$
since $x_i$ and $Q_i$ are uncorrelated.  
Note also that the arrival positions of drift electrons ($x_i$) are not correlated.
}
\input{EqB3.tex}

Hence the spatial resolution as a function of the drift distance ($z$) can be numerically 
evaluated once the pad pitch ($w$), the diffusion constant ($D$), the pad response
function ($f$), and the effective number of electrons ($N_{\rm eff}$) are given. 
In our case the pad response function is assumed to be a gaussian
\begin{displaymath}
f(\xi) = \frac{1}{\sqrt{2\pi} \sigma_a}
                {\rm exp} \biggl( - \frac{\xi^2}{2\sigma_a^2} \biggr) \;,
\end{displaymath}
with $\sigma_a$ being the charge spread in the GEM stack.

Though Eq.~(\ref{b3}) may appear a little complicated its meaning is quite simple:
the first term is the bias inherent in the charge centroid method (finite pad-pitch term)
while the second term represents the variance around the bias, divided by $N_{\rm eff}$
(diffusion term).
This is shown clearly by rewriting Eq.~(\ref{b3}) as
\begin{displaymath}
\sigma_X^2 (\tilde{x}) \approx \biggl< \sum_j F_j(x) \cdot x_j^*-\tilde{x} \biggr>^2 
+ \frac{1}{N_{\rm eff}}
\biggl< \biggl( \sum_j F_j(x) \cdot x_j^*
- \biggl< \sum_j F_j(x) \cdot x_j^* \biggr> \biggr)^2~ \biggr> \;.
\end{displaymath}

It should be pointed out here that $\sigma_X^2$ depends on the position of
$\tilde{x}$ relative to the corresponding pad center,
and that the beam spot size is usually much larger than the pad pitch.
Therefore unless the incident positions of incoming particles are measured
precisely by external trackers (e.g. by a set of silicon strip detectors)
on an event-by-event basis, $\sigma_X^2(\tilde{x})$ obtained above (Eq.~(\ref{b3}))
has to be averaged over $\tilde{x}$ in a range, say, [-$w$/2, +$w$/2]:
\begin{displaymath}
\sigma_X^2 = \frac{1}{w} \cdot \int_{-w/2}^{+w/2} \sigma_X^2(\tilde{x}) \; d\tilde{x} \;.
\end{displaymath}

In order to see the asymptotic behavior of the resolution it is convenient to rewrite
Eq.~(\ref{b3}):
\begin{displaymath} 
\sigma_X^2 (\tilde{x}) \approx \biggl< \sum_j F_j(x) \cdot x_j^*-\tilde{x} \biggr>^2  
+ \frac{1}{N_{\rm eff}}
\biggr( \biggl< \biggl( \sum_j F_j(x) \cdot x_j^* - \tilde{x} \biggr)^2 \biggr>
- \biggl< \sum_j F_j(x) \cdot x_j^* - \tilde{x} \biggr>^2 \biggr) \;,
\end{displaymath}
and define
\begin{equation}
\sigma_X^2 = \frac{1}{w} \cdot \int_{-w/2}^{w/2} \sigma_X^2(\tilde{x}) \; d\tilde{x} 
  \approx B^2 + \frac{1}{N_{\rm eff}} \cdot (A^2 - B^2) \;, 
 \label{b4}
\end{equation}
where
\begin{eqnarray*}
B^2 &\equiv& \frac{1}{w} \cdot \int_{-w/2}^{w/2}
  \biggl< \sum_j F_j(x) \cdot x_j^*-\tilde{x} \biggr>^2 \;d\tilde{x} \;, \\
A^2 &\equiv& \frac{1}{w} \cdot \int_{-w/2}^{w/2}
  \biggl< \biggl( \sum_j F_j(x) \cdot x_j^* - \tilde{x} \biggr)^2 \biggr> \;d\tilde{x} \;.
\end{eqnarray*}
The term $B^2$ vanishes in the limit $\sigma_d \rightarrow \infty\; (z \rightarrow \infty)$
since $\sum_j \left< F_j(x) \right> x_j^* \rightarrow \tilde{x}$ for any given $\tilde{x}$
while the term $A^2$ can easily be shown to be
$B_0^2 + \sigma_d^2 = B_0^2 + D^2 \cdot z$,
where $B_0^2 \equiv B^2\;]_{\sigma_d = 0} = B^2\;]_{z = 0}$.
Therefore 
\begin{equation}
\sigma_X^2 \rightarrow \frac{1}{N_{\rm eff}} \cdot \bigl( B_0^2 + D^2 \cdot z \bigr)
 \label{b5}
\end{equation}
at long drift distances.

\begin{figure}[htbp]
\begin{center}
\includegraphics*[scale=0.60]{\figdir/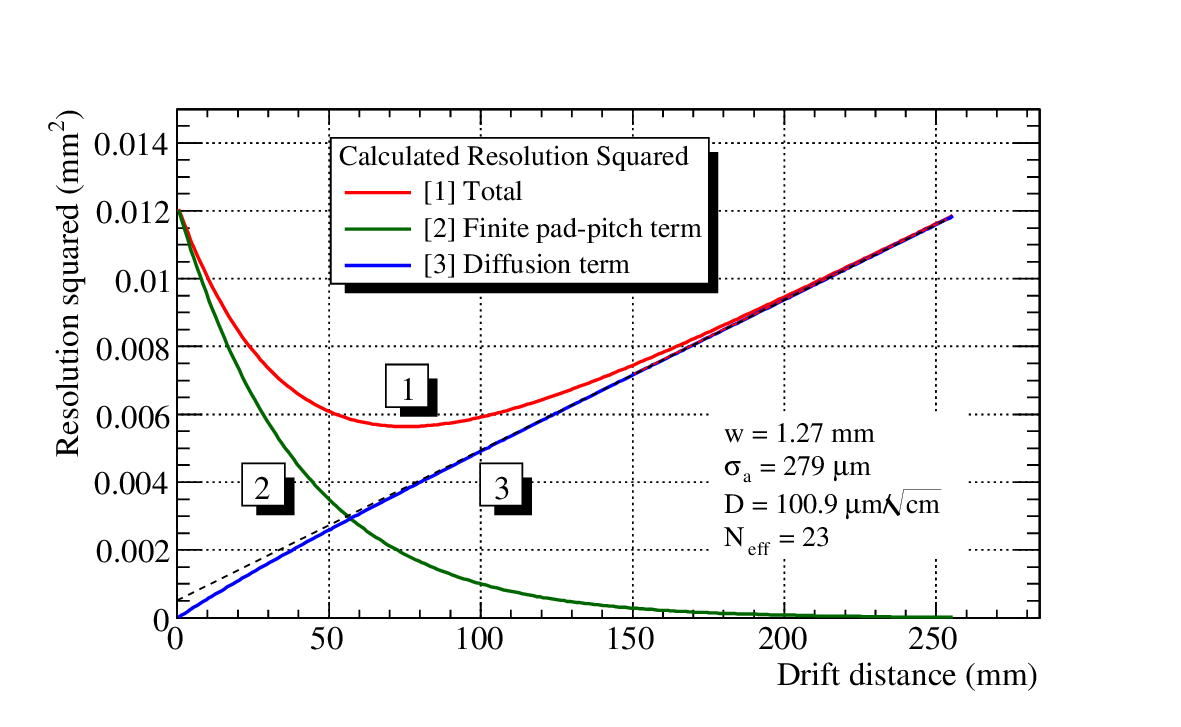}
\end{center}
\caption[figA2]{\label{eee}
\footnotesize Calculated resolution squared for $E_{\rm d}$ = 250 V/cm and $B$ = 1 T. 
See text for the dashed line.
}
\end{figure}

Figure~\ref{eee} shows an example of the calculated resolution in our case.
In the calculation the following values were assumed:
$w$ = 1.27 mm, 
$\sigma_a$ = 279 $\mu$m,
$D$ = 100.9 $\mu$m/$\sqrt{\rm cm}$ (Magboltz value),
and $N_{\rm eff}$ = 23.
$\sigma_a$ was estimated using the diffusion constant in the transfer and the induction gaps
given by Magboltz.
The dashed straight line shows the asymptotic behavior at long drift distances (Eq.~(\ref{b5})),
which has an offset of $B_0^2 / N_{\rm eff} \approx (23\; \mu {\rm m})^2$.
This is the fundamental lower limit of the constant contribution to the spatial resolution
($\sigma_{\rm X0}^2$), determined by
the pad pitch ($w$), the charge spread in the detection gap ($\sigma_a$),
and the effective number of electrons ($N_{\rm eff}$). 
 
%% file: EqB3.tex
\begin{eqnarray}
\sigma_X^2(\tilde{x}) &=& \bigl<(X-\tilde{x})^2 \bigr > \nonumber \\
&=& \biggl< \biggl(
        \frac{\sum_{i=1}^N Q_i \sum_j F_j(x_i) \cdot x_j^*}{\sum_{i=1}^N Q_i}
              - \tilde{x} \biggr)^2 \biggr> \nonumber \\
&=& \biggl< \biggl(
        \frac{\sum_{i=1}^N Q_i \sum_j F_j(x_i) \cdot (x_j^*-\tilde{x})}
                                           {\sum_{i=1}^N Q_i}
                                                \biggr)^2 \biggr> \nonumber \\
&=& \biggl< 
     \frac{1}{(\sum_i Q_i)^2} \cdot 
       \biggl(
         \sum_i Q_i^2 \cdot
           \biggl(
\sum_j F_j(x_i) \cdot (x_j^*-\tilde{x}) \biggr)^2 \nonumber \\ 
&& ~~~~~~~~~~~~~~~~~~~~+
\sum_{i\neq l} Q_iQ_l \cdot
           \biggl(
\sum_k F_k(x_i) \cdot (x_k^*-\tilde{x}) \sum_k F_k(x_l) \cdot (x_k^*-\tilde{x})
           \biggr) 
                   \biggr) \biggr > \nonumber \\
&=& \biggl< 
     \frac{1}{(\sum_i Q_i)^2} \cdot 
       \biggl(
         \sum_i Q_i^2 \cdot
\Bigl( \sum_j F_j(x_i) \cdot x_j^*-\tilde{x} \Bigr)^2 \nonumber \\ 
&& ~~~~~~~~~~~~~~~~~~~~+
\sum_{i\neq l} Q_iQ_l \cdot
           \biggl(
\Bigl( \sum_k F_k(x_i) \cdot x_k^*-\tilde{x} \Bigr)
                  \Bigl( \sum_k F_k(x_l) \cdot x_k^*-\tilde{x} \Bigr)
           \biggr) 
                   \biggr) \biggr > \nonumber \\
&=& \biggl< 
     \frac{1}{(\sum_i Q_i)^2} \cdot 
       \biggl(
         \sum_i Q_i^2 \cdot
           \biggl<
\Bigl( \sum_j F_j(x) \cdot x_j^*-\tilde{x} \Bigr)^2 \biggr>_x \nonumber \\ 
&& ~~~~~~~~~~~~~~~~~~~~~~~~~~~~~~~~~~~ +
\bigl( \sum_{i,l} Q_iQ_l - \sum_i Q_i^2 \bigr) \cdot
           \biggl<
    \sum_k F_k(x) \cdot x_k^*-\tilde{x}
           \biggr>_x^2~ 
                   \biggr) \biggr >_Q \nonumber \\
&=& \biggl< \sum_j F_j(x) \cdot x_j^*-\tilde{x} \biggr>_x^2 
+ \biggl< \frac{\sum_i Q_i^2}{(\sum_i Q_i)^2} \biggr>_Q \cdot
\biggl( \biggl< \sum_{j,k} F_j(x) \cdot F_k(x) \cdot x_j^* \cdot x_k^* \biggr>_x
- \biggl< \sum_j F_j(x) \cdot x_j^* \biggr>_x^2~ \biggr) \nonumber \\
&=& \biggl( \sum_j \bigl< F_j(x) \bigr> \cdot x_j^*-\tilde{x} \biggr)^2 
+ \biggl< \frac{\sum_i Q_i^2}{(\sum_i Q_i)^2} \biggr> \cdot
\biggl(  \sum_{j,k} \bigl< F_j(x) \cdot F_k(x) \bigr> \cdot x_j^* \cdot x_k^*
- \biggl(  \sum_j \bigl< F_j(x) \bigr> \cdot x_j^* \biggr)^2~ \biggr) \nonumber \\
&\approx& \biggl( \sum_j \bigl< F_j(x) \bigr> \cdot x_j^*-\tilde{x} \biggr)^2 
+ \frac{1}{N} \cdot \frac{\bigl< Q^2 \bigr>}{\bigl< Q \bigr>^2} \cdot
\biggl(  \sum_{j,k} \bigl< F_j(x) \cdot F_k(x) \bigr> \cdot x_j^* \cdot x_k^*
- \biggl(  \sum_j \bigl< F_j(x) \bigr> \cdot x_j^* \biggr)^2~ \biggr) \;. \nonumber
\end{eqnarray}
In the last line $\sum_i Q_i$ is approximated by $N \cdot \left< Q \right>$,
assuming large $N$.
Averaging over $N$ and substituting $j \cdot w$ and
$k \cdot w$, respectively for $x_j^*$ and $x_k^*$ we get 
\begin{equation}
\sigma_X^2 (\tilde{x})\approx \biggl( \sum_j jw \cdot \bigl< F_j(x) \bigr>-\tilde{x} \biggr)^2 
+ \frac{1}{N_{\rm eff}} \cdot
\biggl(  \sum_{j,k} jkw^2 \cdot \bigl< F_j(x) \cdot F_k(x) \bigr> 
- \biggl(  \sum_j jw \cdot \bigl< F_j(x) \bigr> \biggr)^2~ \biggr) \;, \label{b3} \\
\end{equation}
where
\begin{eqnarray*}
&& \frac{1}{N_{\rm eff}} \equiv 
  \biggl< \frac{1}{N} \biggr> \cdot \frac{\bigl< Q^2 \bigr>}{\bigl< Q \bigr>^2}
  = \biggl< \frac{1}{N} \biggr> \cdot \biggl( 1 + \frac{\sigma_Q^2}{\bigl< Q \bigr>^2} \biggr) \;,
                                         \nonumber \\ 
&& \bigl< F_j(x) \bigr > \equiv \int_{-\infty}^{\infty} P_x(x) \cdot F_j(x) \; dx \;,
 \hspace{90mm} \nonumber \\
&& \bigl< F_j(x) \cdot F_k(x) \bigr > \equiv
             \int_{-\infty}^{\infty} P_x(x) \cdot F_j(x) \cdot F_k(x) \; dx \;,\nonumber \\
&& {\rm with} \nonumber \\ 
&& P_x(x) \equiv \frac{1}{\sqrt{2\pi} \sigma_d}
                {\rm exp} \biggl( - \frac{(x-\tilde{x})^2}{2\sigma_d^2} \biggr) \;,\nonumber \\
&& F_j(x) \equiv \int_{jw-w/2}^{jw+w/2} f(\xi-x)\, d\xi \;, \nonumber \\
&& f(\xi) : {\rm ~PRF} \nonumber \; .
\end{eqnarray*}
The parameter $\sigma_d$ in the definition of $P_x(x)$ denotes the (lateral) standard deviation of 
the arrival position of drift electrons at the entrance to the amplification device
measured from $\tilde{x}$, and is given by $D \cdot \sqrt{z}$.

%% file: AppendixC.tex
In this appendix we qualitatively discuss the origin of the constant term
contained in the spatial resolution ($\sigma_{\rm X0}^2$) in our case.
Possible main sources of the constant term are listed below
for tracks perpendicular to the pad-row direction.
\begin{enumerate}
\item Intrinsic track width ($\delta$-rays) \cite{Sauli1978}.
\item Multiple Coulomb scattering.
\item Mechanical imprecision.
\item Influence of the finite hole pitch of the GEM foils.
\item Influence of the finite pad pitch.
\item Electronic noise.
\item Inadequate calibration of the readout electronics,
      such as channel-to-channel (pad-to-pad) inequalities of the gain
      and inappropriate pedestal subtractions from the raw signals.
\end{enumerate}
The sources (1) and (2) are intrinsic and unavoidable while the others
can be made sufficiently small in principle.

The intrinsic track width certainly contributes to the constant term.
The value of $\sigma_{\rm X0}^2$ obtained with the magnetic field
($\sim$ (50 $\mu$m)$^2$) is significantly smaller than those
for the resolutions obtained without the magnetic field
($\sim$ (100 $\mu$m)$^2$, see Figs.~13 and 14 for example)
because of the suppression of the transverse intrinsic track width by the
axial magnetic field.
The magnetic field reduces the {\it effective range\/} of $\delta$-rays
along the pad-row direction.

The mechanical precision such as the readout pad arrangement is expected
to be good enough.
The influence of the hole pitch of the GEM foils 
(horizontal pitch of 140 $\mu$m) was also found to be small
by means of a Monte-Carlo simulation
(see, for example, Fig.~4 of Ref.~\cite{Kobayashi2}).
The constant term (at long drift distances) is caused by the finite
pad pitch as well (see Appendix B, Eq.~(\ref{b5})). 
However, its contribution is small in our case,
(18 $\mu$m)$^2$ - (23 $\mu$m)$^2$, depending on the magnetic field
(0 T - 1 T). 

The contribution of the angular pad effect for the azimuthal angle cut of
$\pm$ 3$^\circ$ was estimated to be $\sim$ (20 $\mu$m)$^2$ by
a Monte Carlo simulation.
Therefore the main contributors to $\sigma_{\rm X0}^2$ are expected to be
(6) and (7), except for the unavoidable contribution from (1) and (2).

The sources (6) and (7) have contribution increasing with
the drift distance ($z$) as well as a constant term at $z$ = 0
because of the increase of the number of active pads used to determine the
charge centroid, due to diffusion
(see Figs.~9 and 10, and Eq.~(\ref{b2}) for the contribution of electronic noise).
Therefore they can affect the {\it apparent} value of $N_{\rm eff}$ as well.

%% file: addendum.tex
\begin{frontmatter}



\title{
  Addendum to ``Cosmic ray tests of a GEM-based TPC prototype operated in
  Ar-CF4-isobutane gas mixtures''
  [Nucl. Instrum. Methods Phys. Res. A 641 (2011) 37-47]
}

\input{authorlist.tex}
\end{frontmatter}


\setcounter{figure}{0}

\end{comment}

\begin{center}
\appendix{\underline{{\bf Addendum} to be published in Nucl. Instrum. Methods Phys. Res. A}}
\end{center}

%
%
The finite pad-pitch term shown in Fig.~B2 in Appendix~B can be eliminated, if not completely, 
provided that the charge spread in the GEM stack ($\sigma_a$) is not too small compared to the pad pitch ($w$)
and the electronic noise is small enough so that the threshold for the flash ADCs is set low enough.
It corresponds to the first term in Eq.~(B.3) and the following equivalent equations,
that represents the bias inherent in the charge centroid (barycenter) method used to get the
track coordinate at a pad row ($X$).
The average bias ($\left< \Delta X \right>$) is
a function of the true hit coordinate $\tilde{x}$ (measured from the center of the pad with the maximum charge deposit).
It depends also on the drift distance ($z$) since the lateral diffusion of drift electrons ($\sigma_d$)
is expressed as
\begin{equation}
\sigma_d = D \cdot \sqrt{z}
\end{equation}
where $D$ denotes the transverse diffusion constant.

The average bias is the first term of Eq.~(B.3) before squared: 
\begin{equation}
  \left< \Delta X(\tilde{x}) \right> \equiv \left< X \right> - \tilde{x} \equiv \sum_j \; j \cdot w \cdot \left< F_j(x) \right>  - \tilde{x}
\end{equation}
where
\begin{eqnarray}
&& \bigl< F_j(x) \bigr > \equiv \int_{-\infty}^{\infty} P_x(x) \cdot F_j(x) \; dx  \\
&& P_x(x) \equiv \frac{1}{\sqrt{2\pi} \sigma_d}
                {\rm exp} \biggl( - \frac{(x-\tilde{x})^2}{2\sigma_d^2} \biggr) \\
&& F_j(x) \equiv \int_{jw-w/2}^{jw+w/2} f(\xi-x)\, d\xi  \\
&& f(\xi) \equiv \frac{1}{\sqrt{2\pi} \sigma_a}
                {\rm exp} \biggl( - \frac{\xi^2}{2\sigma_a^2} \biggr)
\end{eqnarray}
with $j$ ($w$), the pad number (pad pitch).

The value of $\left< F_j(x) \right>$ can be calculated for a given $\tilde{x}$ if $\sigma_a$, $D$, and $z$ (drift time) are known.
One can thus obtain a one-to-one correspondence between $\left< X \right>$ and $\tilde{x}$.
Therefore, the value of $X$ given by the charge centroid method can be projected onto a
bias-free estimate of $\tilde{x}$, assuming $X = \left< X \right>$.
Although the projection is not perfect because of fluctuations of $X$ around $\left< X \right>$,
it is the simplest way of removing the bias without relying on the tracking information provided by the other pad rows.

Fig. 1(a) shows the relation between $\tilde{x}$ and $\left< X \right>$.
Fig. 1(b) shows the average bias ($\left< \Delta X \right>$) given by Eq.~(2) as a function of
$\tilde{x}$ measured from the pad center,
while Fig. 1(c) is the spatial resolution, after the bias subtraction, as a function of $\tilde{x}$,
i.e. the square root of the second term of Eq.~(B.3).

Fig.~1(c) shows that the charge centroid after the {\it perfect\/} bias subtraction always gives
the true coordinate ($\tilde{x}$) at $z$ = 0.
The second term in Eq.~(B.3) vanishes at $z$ = 0 for any $\tilde{x}$
because of the assumptions (3)--(5) listed at the beginning of Appendix B.
In the case of GEMs, the item~(4) means an infinitesimal hole pitch.
Consequently, the second term of Eq.~(B.4) (diffusion term) vanishes at $z$ = 0, as shown in Fig.~B2.
The diffusion term is the square of such as that shown in Fig.~1(c) averaged over $\tilde{x}$
for the corresponding drift distance ($z$).

It should be noted that the fluctuations of the charge centroid of multiplied electrons in the GEM stack
for a single drift electron are small when the effective gain of the first GEM is large enough.
Their contribution to the diffusion term is further reduced by the factor $N_{\rm eff}$.

It should also be noted that the contribution of electronic noise (Eq.~(B.2)) is neglected
assuming a large gas gain in the GEM stack.
See Appendix C for other possible sources of the degradation of the spatial resolution.

By definition, the {\it diffusion\/} term depends on the pad pitch ($w$) as well as the drift distance ($z$).
It is, therefore, appropriate to call the first (second) term of Eq.~(B.4) {\it bias\/} ({\it statistical fluctuation\/}) term
independent of (dependent on) $N_{\rm eff}$.
\begin{figure}[htbp]
\begin{center}
\includegraphics*[scale=0.44]{\figdir/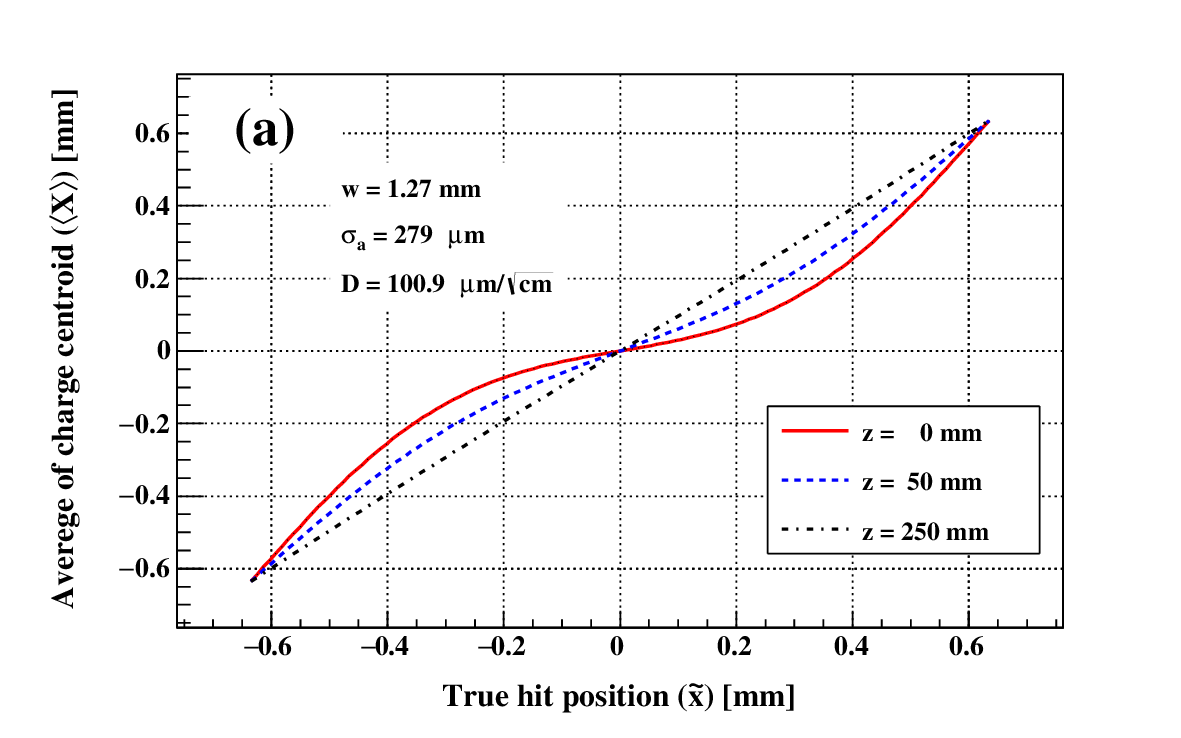}
\includegraphics*[scale=0.44]{\figdir/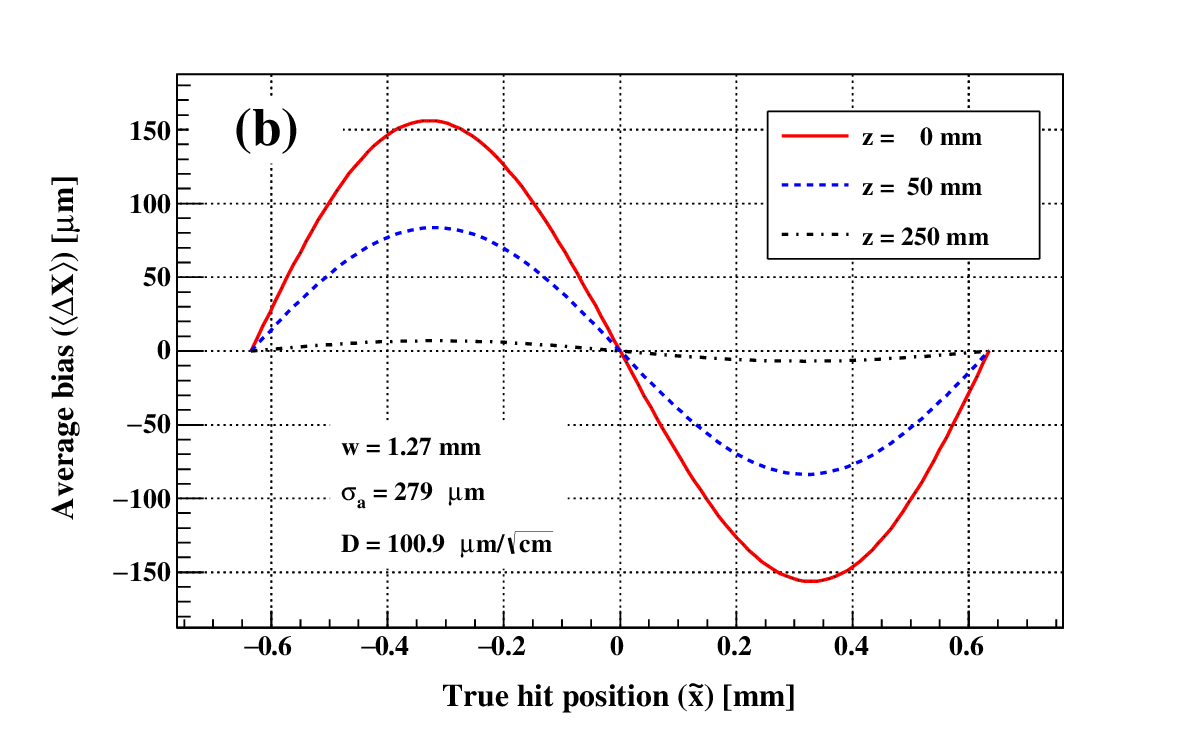}
\includegraphics*[scale=0.44]{\figdir/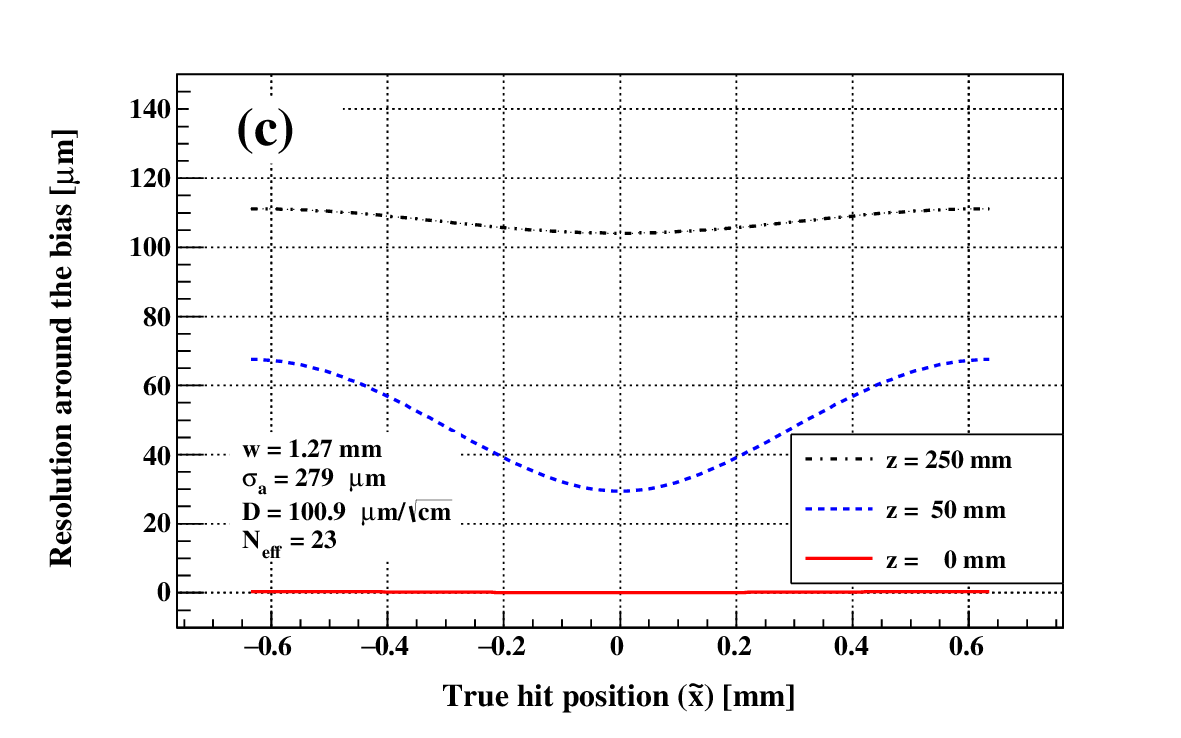}
\caption{\label{fig1}
  \footnotesize (a) Relation between the average of the charge centroid ($\left< X \right>$) and the true hit coordinate
   measured from the pad center ($\tilde{x}$).
   (b) Average bias ($\left< \Delta X \right>$) as a function of $\tilde{x}$.
   (c) Spatial resolution after bias subtraction as a function of $\tilde{x}$.
}
\end{center}
\end{figure}
